\documentstyle[aps,prc,manuscript]{revtex} 
\include{psfig}
 
\begin{document} 
\tighten 
\draft 
 
\title{Neutrino Capture and r-Process Nucleosynthesis} 
\author{Bradley S. Meyer} 
\address{Department of Physics and Astronomy, Clemson University, 
Clemson, SC 29634-1911} 
\author{Gail C. McLaughlin} 
\address{University of Washington, Institute for Nuclear Theory,
Box 351550, Seattle, WA 98195-1550} 
\author{George M. Fuller} 
\address{Department of Physics, University of California, San Diego,
La Jolla, CA 92093-0319} 
\date{\today} 
\maketitle 

\begin{abstract}
We explore neutrino capture during r-process nucleosynthesis in
neutrino-driven ejecta from nascent neutron stars.  We focus on
the interplay between charged-current weak interactions and
element synthesis, and we delineate the important role of equilibrium
nuclear dynamics.
During the period of coexistence of free nucleons 
and light and/or heavy nuclei, electron neutrino capture inhibits 
the r-process.  At all stages, capture on free neutrons has a larger 
impact than capture on nuclei.  However, neutrino
capture on heavy nuclei by itself,
if it is very strong, is also detrimental to the
r-process until large nuclear equilibrium clusters break down and the
classical neutron-capture phase of the r-process begins.  The
sensitivity of the r-process to neutrino irradiation means that 
neutrino-capture effects can
strongly constrain the r-process site, neutrino physics, or both.
These results apply also to r-process scenarios other than the
neutrino-heated winds. 
\end{abstract}
 
\section{Introduction} 

It has long been known that the r-process of nucleosynthesis is responsible
for roughly half the solar system's supply of heavy nuclei
\cite{B2FH,cow91}.  Nevertheless, the astrophysical site or sites
of the r-process remain a great mystery.  The high neutron densities
and rapid timescales associated with the r-process
suggest core-collapse (Type II or Type Ib)
supernovae as the most likely setting, but the exact environment
within supernovae is unclear.  The most plausible environment yet proposed
is the neutrino-heated ejecta from the nascent neutron
star\cite{wh92,mey92,how93,taka94,woo94}.  Neutrinos from the
Kelvin-Helmholz-cooling neutron star heat matter strongly.  Given
sufficient heating, this material can escape the deep gravitational well
and travel into interstellar space along with the rest of the stellar
ejecta.  Necessarily, this neutrino heating drives the entropy per
nucleon to a large value ($\sim 100k$, where $k$ is Boltzmann's constant).
Crucially for the r-process, the emerging electron antineutrinos come from
deeper in the neutron star than the electron neutrinos.  This results from
the larger opacity of the latter in the interior of the neutron star.  As
the neutrinos and antineutrinos capture on free neutrons and protons in the
heated ejecta, the hotter antineutrinos drive the matter neutron rich 
\cite{qian93}.  The
high entropies, fast expansion, and neutron richness of the ejecta may provide
the right conditions for making r-process nuclei.
However, present supernova models with standard neutrino physics do not
attain the extreme conditions needed to make the heaviest r-process
isotopes.
The necessary 
conditions conceivably could be realized by invoking general relativistic 
effects,
though these models are finely tuned at best \cite{cardall97,qw97}.  As we
will argue, however, even if these necessary conditions could be attained,
they are not {\it sufficient} to guarantee a viable r-process.

While neutrino interactions with nuclei are generally not an important
effect in stellar nucleosynthesis (apart from normal beta decay and electron
capture), neutrinos so completely dominate the environment just outside a
newly born neutron star that their effects must be included in
nucleosynthesis calculations done in the context of neutrino-driven ejecta.
Initially, the electron fraction above the surface of the neutron star is set 
primarily by electron neutrino and 
electron antineutrino capture on free nucleons.  Other neutrino capture
effects occur, however, and have been studied.
Meyer et al. \cite{mey92} included
neutral-current spallation of neutrons from nuclei in one of their models.
They found some smoothing of the resulting r-process distribution, but
the overall effect on the r-process yields was small. This was also studied
by Qian et al. \cite{qvh97}.  Meyer \cite{mey95}
then showed that spallation of abundant $^{4}$He in wind trajectories
studied by Woosley et al. \cite{woo94} had a large, detrimental effect on
the r-process.  Fuller and Meyer \cite{fullermeyer95} and McLaughlin,
Fuller, and Wilson \cite{mfw96} studied
charge-current interactions on free nucleons and nuclei during expansions
of neutrino-heated ejecta.  The key finding in those works was the strong
``alpha effect'' in which the electron fraction grows rapidly as $^4$He
nuclei assemble in the presence of a large neutrino flux.  We verify this 
effect in Sec. \ref{sec:results}, and show that of all neutrino effects, it
has the largest impact on the r-process yields.

Neutrino capture on heavy nuclei in competition with nuclear beta decay 
provided limits on supernova dynamics from the r-process, although 
significant capture on the mass number ${\rm A} = 130$
peak was not necessarily excluded \cite{fullermeyer95,mf97}.  However, 
establishing steady weak flow (the analogue 
of steady beta flow) would require a long timescale ($> 1$s) and 
therefore many 
neutrino captures \cite{mf97}.  Following on these studies and work by 
Nadyozhin and Panov\cite{nd93},
Qian et al.  \cite{qvh97} proposed that neutrino capture was needed to
accelerate the r-process.  The basic idea is that neutrino capture acts
like beta decay in moving nuclei to higher charge.  Faster movement upward
in charge is a faster r-process.  This scenario 
would also require many neutrino captures. 

We show in this paper that if neutrino capture on nuclei is significant
compared to nuclear beta decay, then other effects will destroy the
possibility of any r-processing at all.  The basic reason for this is
simple: neutrino capture on a free neutron is much faster than on a
neutron bound inside a nucleus because the former has many more final
states available.  For example, the neutrino capture cross
section of black body neutrinos with temperature $T_{\nu_e} = 3.5$ MeV on a
free neutron is about $1.6 \times 10^{-41}\ {\rm cm}^2$.  The cross section
for the same neutrinos on a neutron bound inside $^{130}$Cd, a typical
r-process nuclide, is about $8.6 \times 10^{-43}\ {\rm cm}^2$.  During the
r-process, the abundance of neutrons inside and outside nuclei are
comparable; thus, capture of a neutrino by a heavy nucleus  will be
accompanied by $( \sim 5-10)$ captures on free neutrons.  This strongly depletes
the supply of free neutrons and limits the r-process.  This process is
similar to the alpha effect and is examined in Sec. \ref{sec:results}.
Furthermore, as we
show in Sec. \ref{sec:heavy}, each neutrino capture on a free neutron
leads to assembly of new heavy nuclei, further depleting the supply
of free neutrons per seed nucleus.

In addition to these considerations, we show in Sec. \ref{sec:qse} that
neutrino capture on heavy nuclei {\em alone}, if it is very strong, does
not accelerate the r-process but rather strongly limits it.  This
surprising result comes about because of the quasi-equilibrium
nuclear dynamics in the early part of the matter expansion.  We describe
these nuclear dynamics in Sec. \ref{sec:hier}.  A proper understanding of
these effects is important for any who seek to use the r-process to
constrain supernova neutrinos.

The strong sensitivity of the r-process to neutrino irradiation presents a
great challenge.  In particular, the effects we explore here constrain
either the r-process site or neutrino physics.  We present some remarks on
the implications of our work in Sec. \ref{sec:implications}.  Preliminary
results were presented in \cite{meyermon97}.
 
\section{Neutrino-Capture Cross Sections} \label{sec:nucap} 

Electron neutrino and antineutrino capture on free neutrons ($n$) and
protons ($p$),
\begin{equation}
\label{eq:ccf}
\nu_e + {\rm n}  \rightarrow {\rm p}+ e^{-};
\end{equation} 
\begin{equation}
\label{eq:ccbf}
\bar{\nu}_e + {\rm p} \rightarrow {\rm n} + e^{+}.
\end{equation}
play an important role in an r-process environment with a 
large neutrino flux. In a supernova, antineutrinos have
a higher average energy and a higher luminosity than neutrinos, causing the 
material in the neutrino-driven wind to be neutron rich. In the following 
sections we discuss additional effects 
which arise from the process in Eq. (\ref{eq:ccf}). 
We also consider effects stemming from 
electron neutrino and antineutrino capture on heavy nuclei,
\begin{equation}
\label{eq:cc}
\nu_e + ({\rm Z},{\rm A}) \rightarrow ({\rm Z+1},{\rm A}) + e^{-};
\end{equation} 
\begin{equation}
\label{eq:ccb}
\bar{\nu}_e + ({\rm Z},{\rm A}) \rightarrow ({\rm Z-1},{\rm A}) + e^{+},
\end{equation}
where $Z$ is the nuclear charge and $A$ the nuclear mass of species
$(Z,A)$.
Since the material undergoing synthesis is neutron rich, in general 
Eq. (\ref{eq:cc}) has a larger impact than Eq. (\ref{eq:ccb}).
In this section we describe our calculations of 
the cross sections for these neutrino processes.  We also discuss our
estimates of neutrino-induced neutron emission processes, which have been 
included in the network calculations.
In calculating all capture rates, we
assume that the distribution of neutrinos coming from the proto-neutron
star is Fermi-Dirac with zero chemical potential.  Though the actual
neutrino and antineutrino distribution functions will deviate somewhat from
our assumed spectral form, this will not change our qualitative conclusions.

For typical neutrino energies in the supernova,
the nuclear channels most important for
electron neutrino and antineutrino capture 
are the allowed Fermi and Gamow-Teller transitions.  
The corresponding operators are 
\begin{equation}
|M_{\rm F}|^2 = | \langle \psi_f | C_V \sum_{i=1}^{A}
{ \bf \tau}(i) | \psi_i \rangle |^2
\end{equation}
\begin{equation}
|M_{\rm GT}|^2 = | \langle \psi_f | C_A 
\sum_{i=1}^{A} {\bf \sigma}(i) { \bf \tau}(i) | \psi_i
\rangle |^2,
\end{equation}
where the sums run over each of the $A$ nucleons in the nucleus.
The Gamow-Teller strength obeys the Ikeda sum rule, 
$ S_{\beta^-} - S_{\beta^+} = 3 |N-Z|$.  For neutron-rich nuclei,
$S_{\beta^-}$ is the Gamow-Teller strength in the neutrino-capture 
direction, while
$S_{\beta^+}$ corresponds to the antineutrino capture
direction.  For very neutron-rich nuclei, the antineutrino capture 
direction is Pauli blocked, $S_{\beta^+} = 0$, 
so the corresponding transition rate is zero.  In these cases, the rate of
antineutrino capture is negligible in comparison with neutrino capture
since there is no Fermi resonance in the $\beta^+$ direction.

In order to fully determine the effect of neutrino interactions 
during the r-process we need rates, cross sections and
particle emission probabilities for a wide range of nuclei.  Although
it would be desirable to use shell model and continuum random phase
approximation (CRPA)
calculations for each nucleus, this is clearly impossible at present.
Therefore, we adopt a simpler, more feasible approach using the single 
particle shell model.  Where possible, we have verified that our results 
agree within reasonable errors with more detailed calculations.
   
The matrix elements and capture rates
are calculated as described in \cite{fullermeyer95,mf95}, 
with some improvements. 
Most of the Fermi strength and Gamow-Teller strength is 
often collected in resonances.  The Fermi resonance is
narrowly collected in a single state, while the Gamow-Teller
resonance has a wider distribution.  We have modified our calculation
to account for this width.  This alteration has a 
relatively small impact on the overall transition
rates (see \cite{mf95}), but can have a slightly larger impact on 
particle spallation, which is described below.  Another modification is
the inclusion of a full integration of the Coulomb wave correction 
factor in the phase space integral.  Our network has also been extended to
include proton-rich nuclei, although in a robust r-process, relatively few
of these appear. In a few cases where comparison is possible, 
our results are in reasonable agreement with calculations done 
using CRPA \cite{qvh97}.  

In addition we have included neutrino-induced neutron emission processes.  
For the very neutron rich
nuclei typical of the r-process, several neutrons will be 
emitted after neutrino-induced excitations to the
Gamow-Teller and Fermi resonances.  We calculate the number of neutrons 
emitted by comparing the neutron separation energy plus kinetic energy 
of each emitted neutron to the excitation energy of the nucleus.  
An estimate of the kinetic energy of an emitted neutron is:
\begin{equation}
\langle E \rangle \approx {\int^{E^* - S_n}_{0} e^2 \rho(E^* - S_n - e)de
\over \int^{E^* - S_n}_{0} e \rho(E^* - S_n - e)de},
\end{equation}
where the level density is 
$\rho(e) \propto \exp[2 (a e)^{1/2}]$ in the Fermi gas model.
Here, $E^*$ is the excitation energy of the nucleus, $S_n$ is the neutron 
separation energy, and $a$ is the Fermi gas constant.  For multiple neutron 
emission, we examine the excited
state and neutron separation energy
of each successive daughter nucleus when calculating the average 
kinetic energy carried away by the neutrons.  We adopt this approach over a 
wide range of nuclei. We have 
compared our results to the neutrino-induced 
neutron emission probabilities calculated with the statistical model
\cite{qvh97} and found them to be in reasonable agreement.

\section{Details of the Fluid Dynamics and Network Calculations} 
\label{sec:dynamics}
Our primary interest is to study the neutrino-capture effects in 
neutrino-heated ejecta from nascent neutron stars.  The relevant fluid 
trajectories are then neutrino-driven winds.  Such winds have been studied 
in several papers (e.g. \cite{taka94,woo94,qw97,duncan86}).  A simple
estimate of the wind parameters can be obtained by 
assuming a constant entropy and radiation-dominated
outflow where the enthalpy per baryon is equated to the gravitational
potential per baryon.  With these assumptions and the additional
assumption that the mass outflow rate ($\dot M = 4 \pi r^2 \rho v$) 
is constant with time, then
it can be shown that 
the ejected matter expands homologously such that 
the radial outward velocity $v$  is 
\begin{equation} 
v \propto r,                               \label{eq:homologous} 
\end{equation} 
where $r$ is the radial coordinate.  Solution of eq. (\ref{eq:homologous}) 
yields 
\begin{equation} 
r = r_0 \exp({t/\tau}),                      \label{eq:rad} 
\end{equation} 
where $r_0$ is the initial radius and $\tau$ is the constant expansion 
timescale 
\begin{equation} 
\tau = r / v. 
\end{equation} 
With the above assumption 
of constant entropy and an assumed  $\propto 1/r$ scaling
for the enthalpy per baryon, 
the density $\rho$ scales as 
\begin{equation} 
\rho \propto r^{-3}.                      \label{eq:density} 
\end{equation}
This follows because an equation of state completely dominated 
by relativistic particles has an entropy per baryon,
\begin{equation} 
S \approx {2 \pi^2 \over 45} {1 \over N_A} g_s {T^3 \over \rho},
 \label{eq:tpropr} 
\end{equation} 
where $g_s$ is the effective relativistic particle statistical weight.
In our numerical calculations, we include not only the relativistic photons 
in our equation of state 
but also the partially (in general) relativistic $e^+-e^-$ pairs and the 
non-relativistic nuclear species where appropriate.  
Including nonrelativistic degrees of freedom in a reckoning of the
entropy results in a 
somewhat different behavior for $T$ than that given in Eq. (\ref{eq:tpropr}) 
(see below). 
 
We performed our calculations with the Clemson nucleosynthesis code 
\cite{mey95,mey96,mey98a}.  This is a fully implicit, single-network code 
that includes over 3000 nuclear species ranging from neutrons and protons to 
actinide nuclei.  The reaction network used for the present calculation 
included isotopes for each element 
from the proton-drip to the neutron-drip lines.  We used neutrino-capture 
rates computed as in Sec. \ref{sec:nucap}.  The neutrinos were taken to have 
Fermi-Dirac energy spectra with zero chemical potential and
with temperatures $kT_{\nu_e} = 3.5$ MeV and 
$kT_{\bar{\nu}_e} = 4.0$ MeV.  Because we are not studying neutral-current 
effects in this paper, we ignored the $\mu$ and $\tau$ neutrinos.  
The luminosity in 
electron neutrinos was taken to be
$10^{51}$ ergs/s while that of the electron
anti-neutrinos was taken to be $4 \times 10^{51}$ ergs/s.  The 
large discrepancy in the electron neutrino and anti-neutrino luminosities 
was necessary to provide ejecta that were neutron rich enough to guarantee
the {\it necessary} conditions for r-process nucleosynthesis to occur.  
 
All of our calculations were at performed at 
constant entropy.  In the true expansion
out-of-equilibrium nuclear reactions during 
the expansion generate entropy.  Nevertheless, for sufficiently high 
entropy (greater than $\sim 50k$ per nucleon) the generated entropy has 
little effect on the nuclear abundances \cite{meyerbrown97}.  
Our technique was 
the following.  We began with $r_0 = 5.6$ km, an entropy of 110$k$ per 
nucleon, a density of $2\times 10^8$ g/cm$^3$, and an expansion timescale 
$\tau = 0.3$ s.  This corresponded to an initial temperature of 
$T_9 = T/10^9{\rm K} \approx 40$.  
The temperature and density track at this early stage has 
little effect on the resulting nucleosynthesis.  We only extend the
beginning of the calculation to such high temperatures to ensure 
that at $T_9 = 10$ our nuclear reaction calculations
begin with an electron fraction that accurately
represents a steady state between
neutrino and antineutrino capture rates, with a small contribution from
electron and positron capture.  At each timestep $t+\Delta t$, we 
updated the radius $r(t+\Delta t)$ [Eq. (\ref{eq:rad})] and density $\rho(t + 
\Delta t)$ [Eq.  (\ref{eq:density})].  We then estimated the new temperature 
$T(t+\Delta t)$ and updated the abundances $Y(t+\Delta t)$ with the nuclear 
network.  We next computed the entropy by inverting the relevant integral
equations for the electron fraction and the entropy \cite{meyerbrown97}
and compared the new entropy at $t+\Delta t$ with the old entropy at 
$t$.  If the entropies at $t$ and $t+\Delta t$ did not agree, we tried a 
new temperature and repeated the procedure.  Once we found $T(t+\Delta t)$ 
such that the entropies agreed, we moved on to the next timestep. 
 
\begin{figure} 
\centerline{\psfig{figure=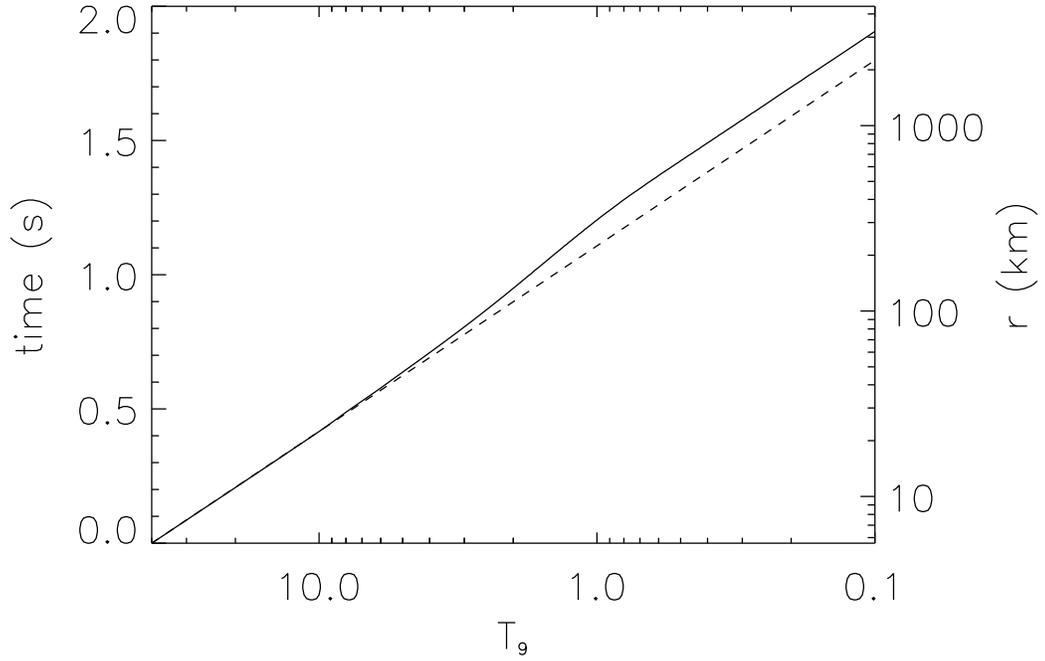,height=4.0in,angle=90}} 
\caption{ 
The temperature-time or temperature-radius trajectory for the reference 
expansion (model 0), shown as the solid line.  The dashed curve gives the 
trajectory for matter completely dominated by relativistic particles, for 
which $\rho \propto T^3$ and $T \propto 1/r$.  The reheating in the 
reference expansion results from the annihilation of electron-positron pairs, 
just as in the early universe. 
} 
\label{fig:rad} 
\end{figure}

\begin{figure} 
\centerline{\psfig{figure=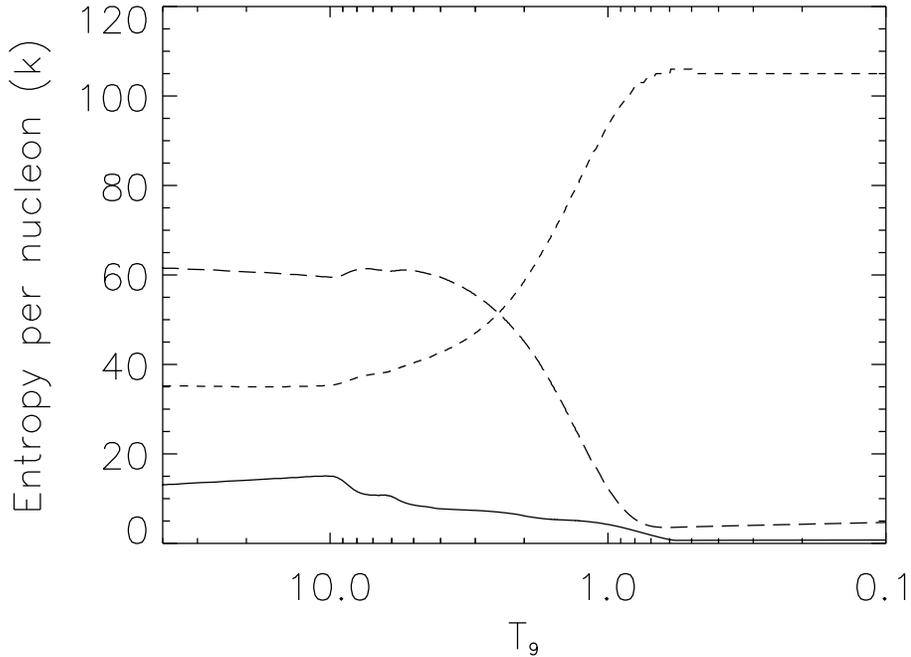,height=4.0in,angle=90}} 
\caption{ 
The entropy per nucleon in units of Boltzmann's constant $k$ in nuclear 
species (solid curve), photons (short-dashed curve), and leptons 
(long-dashed curve) during the reference expansion.  The entropy is a 
constant throughout the expansion, so the constituent entropies always sum 
to 110$k$ per nucleon.  The pair annihilation 
between $T_9 \approx 3$ and $T_9 \approx 1$ shifts entropy from the leptons 
to the photons.  The entropy in nuclear species declines as free nucleons 
lock up into heavier species.  This decreases the number of degrees of 
freedom per nucleon. 
} 
\label{fig:entropy} 
\end{figure}

Figure \ref{fig:rad} shows the trajectory for our reference calculation in 
which neutrinos were ``turned off'' after the matter had cooled below $T_9 
= 10$.  The solid line gives the actual temperature-time or 
temperature-radius relation while the dashed line gives the result for a 
purely $\rho \propto T^3$ case.  The actual network calculation does not 
follow the simple $T \propto 1/r$ expected if $\rho \propto T^3$.  Rather, 
there is heating of the matter due to $e^+-e^-$ annihilations from $T_9 
\approx 8 - 0.8$, that is, from about 0.5-1.2 seconds in the expansion. 
This is precisely analogous to the electron-positron annihilation that 
occurred in the early universe.  This is also evident in Fig. 
\ref{fig:entropy}, which shows the entropies in the leptons, photons, and 
nuclear species during the expansion.  The leptons transfer their entropy 
into the photons rather dramatically, especially around $T_9=3-2$, as the 
pairs annihilate.  The decline in entropy in nucleons results from nuclear 
reactions that lock free nucleons into nuclei, thereby reducing the number 
of degrees of freedom per nucleon.  The gradual drop in the entropy in 
nuclear species from $T_9 = 2$ down to $T_9 < 0.2$ signals the classical 
r-process phase in which free neutrons are being incorporated into heavy 
nuclei.  The calculation proceeded until the abundance $Y_n$ of free neutrons 
per baryon dropped to below $10^{-20}$.

With our chosen conditions, final reaction freezeout occurred at $T_9
\approx 0.5$, a radius $r \approx 1000$ km, and a density of about 35 g/cc.
A faster expansion could
lead to freezeout at an even greater radius and smaller density.
A very fast expansion potentially poses problems for the r-process.
As discussed above, wind matter may expand
homologously such that the outflow velocity is proportional to the radial
distance from the neutron star.  In such a case, the density of a wind
element falls exponentially with the time.  This means the density could
be declining so rapidly that the r-process would freeze out before
all the neutrons could be incorporated into nuclei.  The perceived need to
circumvent this freezeout problem by
accelerating the r-process with neutrino capture motivated some of the
previous studies of neutrino capture during the r-process.

The next sections will show
that rapid neutrino capture certainly would solve this
potential problem, but at the rather drastic cost of eliminating the
possibility of the r-process in the first place.  On the other hand, for
the r-process to successfully occur in neutrino-heated ejecta, matter must
travel out sufficiently rapidly to escape the harmful effects of the
neutrinos.  This would appear to call for a very rapid expansion with the
attendant difficulty of freezeout before successful incorporation of
neutrons into nuclei.  The solution to this dilemma is likely to be that
neutrino-driven winds cannot continue to accelerate so
rapidly that the nuclear reactions would freeze out before the r-process
finished.  Neutron capture can occur rapidly even in relatively cold matter
down to densities of order 1-10 g/cc, and it is unlikely that neutron star
matter would be able to expand homologously from neutron star densities
down to this low value.  At some distance the acceleration provided by 
the neutrinos declines and the matter from then on 
would travel out with a more or less constant
velocity.  In this case the density declines only as the inverse square of
the time for constant mass loss.
This is much slower and would allow the r-process to occur.

\section{The r-Process}   
\label{sec:hier} 
 
Before presenting the details of the 
nucleosynthesis calculations, it is useful to review 
the basics of the r-process.  A proper understanding helps clarify neutrino 
effects. 
 
The nuclear dynamics of matter expanding from high density and 
temperature is probably best viewed as a descent of the hierarchy of 
statistical equilibria \cite{mey98a}. 
Each equilibrium is an entropy maximum 
subject to some number of constraints on the nuclear populations. 
The top of the hierarchy is the 
equilibrium with the fewest constraints.  As the matter expands and cools, 
some nuclear reaction becomes too slow to maintain that equilibrium.  This 
imposes a new constraint on the equilibrium.  With further expansion, other 
reactions become too slow, and new constraints appear.  Reaction freezeout 
occurs when there are the maximum possible number of constraints on the 
nuclear populations.  It is useful to note that the greater the number of 
constraints, the greater the order.  Order emerges in such systems as the 
number of states available to the system increasingly
falls short of the maximum possible. 
 
In the r-process, matter can begin at sufficiently high temperature and 
density that the nuclei are in weak statistical equilibrium (WSE).  Here 
all strong, electromagnetic, and weak interactions among nuclei proceed 
sufficiently rapidly that the only constraints on the equilibrium are that 
charge neutrality holds and that 
the baryon and lepton numbers are fixed, as is the energy if the matter 
comprises an isolated system.  To find the particle abundances in this 
equilibrium, it is only necessary to specify the temperature $T$ and 
density $\rho$.  An entropy maximization via Lagrange multipliers yields
beta equilibrium 
\begin{equation} 
\mu_p + \mu_{e^-} = \mu_n + \mu_{\nu_{e}},        \label{eq:weak1} 
\end{equation} 
where the different $\mu$'s are (total energy) 
chemical potentials for protons, electrons, 
neutrons, and electron-type neutrinos, respectively.  
This equation indicates equilibrium in the interconversion of neutrons and 
protons by the weak interactions induced by electrons  and/or positrons 
and electron neutrinos and /or electron antineutrinos. 
Another result is 
\begin{equation} 
\mu_{e^+} = -\mu_{e^-},                            \label{eq:weak2} 
\end{equation} 
which indicates, for example, electromagnetic equilibrium 
between electrons and positrons.  Finally, the chemical potential for nuclear 
species $(Z,A)$ with charge number $Z$ and mass number $A$ satisfies 
\begin{equation} 
\mu(Z,A) = Z \mu_p + N \mu_n,                 \label{eq:weak3} 
\end{equation} 
where $N=A-Z$ is the neutron number for that species. 
The ideal Boltzmann gas expression for the chemical potential of the 
nuclear species is 
\begin{equation} 
\mu(Z,A) = m(Z,A) c^2 + kT\ln \left( {Y(Z,A) \over Y_Q(Z,A)}\right), 
\label{eq:mudef} 
\end{equation} 
where $m(Z,A)$ is the mass of nuclear species $(Z,A)$, $k$ is Boltzmann's 
constant, $T$ is the temperature, $Y(Z,A)$ is the abundance per nucleon of 
$(Z,A)$, and $Y_Q(Z,A)$ is the quantum abundance of nucleus 
$(Z,A)$ per nucleon. 
$Y_Q(Z,A)$ is given by 
\begin{equation} 
Y_Q(Z,A) = \left({m(Z,A) k T \over 2 \pi \hbar^2} \right )^{3/2} 
{G(Z,A) \over \rho N_A},   \label{eq:yqdef} 
\end{equation} 
where $G(Z,A)$ is the nuclear partition function of $(Z,A)$ and $N_A$ is 
Avogadro's number.  It is useful to note that the quantum concentration
$n_Q(Z,A) = \rho N_A Y_Q(Z,A)$ is the number density associated with
one nucleus $(Z,A)$ in a cube of side roughly equal to the thermal average de
Broglie wavelength of that nucleus.
The resulting nuclear abundances per nucleon are then 
(see \cite{B2FH})
\begin{equation} 
Y(Z,A) = Y_Q(Z,A) \left({Y_p \over Y_{Qp}}\right)^Z \left({Y_n \over 
Y_{Qn}}\right)^N \exp(B(Z,A)/kT),        \label{eq:nseabund} 
\end{equation} 
where the nuclear binding energy $B(Z,A)$ of species $(Z,A)$ is 
\begin{equation} 
B(Z,A) = Zm_pc^2 + Nm_nc^2 - m(Z,A)c^2    \label{eq:bdef} 
\end{equation} 
 
The first reactions to become too slow to maintain full equilibrium during 
the expansion are usually the weak interactions.  First the electron and
positron capture reactions drop out of equilibrium.  Since the electron
neutrino and antineutrino capture reactions drop out later, the electron
fraction is essentially determined by the neutrino reactions.
After the weak reactions become slow, the electron-to-nucleon 
fraction $Y_e$ is not its WSE value and must now be specified.
$Y_e$ 
is given from charge neutrality by 
\begin{equation} 
Y_e = \sum_{Z,A} Z Y(Z,A),                 \label{eq:yedef} 
\end{equation} 
where the sum runs over all nuclear species
including free nucleons.  All other reactions proceed 
rapidly to maintain equilibrium.  This new equilibrium, nuclear statistical 
equilibrium (NSE), is an entropy maximum just like WSE, but now has the 
extra constraint on $Y_e$.  The NSE abundances are given by Eq. 
(\ref{eq:nseabund}), but these must now satisfy Eq. (\ref{eq:yedef}) in 
place of Eq. (\ref{eq:weak1}).  The extra constraint on NSE 
locates it lower in the hierarchy of statistical equilibria than WSE.
It is important to understand that the constraint on $Y_e$ does not mean it is
fixed in time, rather that it changes more slowly than needed to maintain
WSE.  All other reactions are occurring rapidly.
 
The next reactions during the expansion to become too slow are usually the 
three-body reactions that assemble alpha particles into heavier nuclei. 
In the neutron-rich matter required for the r-process, these reactions are 
triple-$\alpha$ ($\alpha + \alpha + \alpha \to ^{12}$C) and the $^{9}$Be 
sequence ($\alpha + \alpha +n \to ^9$Be followed by 
$^9$Be$(\alpha,n)^{12}$C).  Among strong and electromagnetic reactions, 
these can be the slowest because they require three instead of the usual 
two particles to collide. 
The slowness of these reactions keeps the system from maintaining $Y_h$, 
the abundance of heavy nuclei (i.e. those nuclei with $A \geq 12$), at the 
full equilibrium value.  The definition of $Y_h$ is 
\begin{equation} 
Y_h = \sum_{Z,A\geq12} Y(Z,A),         \label{eq:yhdef} 
\end{equation} 
and the slowness of the three-body reactions now constrains $Y_h$ to a 
specified value.  Nevertheless, all other strong and electromagnetic
reactions proceed rapidly.  The 
resulting equilibrium is called quasi-equilibrium (QSE) \cite{bod68}.  In 
it, heavy nuclei are all in equilibrium with each other under the exchange 
of light particles ($n,p,\alpha$), although the number of heavy nuclei 
differs from that in NSE.  The abundances in QSE are given by \cite{mey98a} 
\begin{equation} 
Y^{QSE}(Z,A) = e^{\mu_h/kT} R_p^Z R_n^N Y^{NSE}(Z,A),    \label{eq:qseabund} 
\end{equation} 
where $\mu_h$ is the chemical potential of heavy nuclei (the energy required to 
add a new heavy nucleus at constant entropy), $R_p=Y_p/Y_p^{NSE}$ and 
$R_n=Y_n/Y_n^{NSE}$ are the overabundances of neutrons and protons compared 
to NSE, and $Y^{NSE}(Z,A)$ is the NSE abundance of $(Z,A)$ at the same 
$T$, $\rho$, and $Y_e$ as the QSE.  The QSE abundances must satisfy 
baryon number conservation and Eqs. 
(\ref{eq:yedef}) and (\ref{eq:yhdef}), which give three 
equations to solve for the three unknowns $\mu_h$, $R_p$, and $R_n$.
It is possible to use Eq. (\ref{eq:yedef}) in solving for the QSE or
NSE solutions at a given instant of time 
even though the (out-of-equilibrium) 
$Y_e$ is changing, as long as it is changing
on a timescale slow compared to that for the reactions in equilibrium.
The same is true for Eq. (\ref{eq:yhdef}) and a slowly changing
$Y_h$ in QSE.
 
A clear picture of the QSE aspect of the expansion of r-process matter is 
important for understanding the effects of neutrino capture.  One might 
suppose that neutrino capture on neutron-rich nuclear species simply 
increases the average nuclear charge.  In a QSE, however, the nuclei are 
all interlocked in a large competitive equilibrium, and the abundances 
are set by a Darwinian struggle among the species.  The 
\lq\lq fittest\rq\rq\ species tend to win (i.e. have large abundances), 
and these are nuclei with strong nuclear 
binding.  Neutrino capture increases $Y_e$.  An increase of $Y_e$ leads to
more proton-rich nuclei in QSE.  Protons would be less bound in such
nuclei, so it is possible that the QSE could adjust itself by disintegrating
some protons from nuclei which would thereby lower the average nuclear charge.
It is only after the QSE breaks down that neutrino 
capture would unambiguously increase the average nuclear charge.  We
illustrate these effects in Sec. \ref{sec:qse}. 
 
The large QSE among all heavy nuclei breaks down when certain reactions 
among the heavy species become too slow to maintain the equilibrium.  At 
this point the nuclear system breaks up into smaller QSE clusters.  The 
nuclei within these clusters are in equilibrium under exchange of light 
particles, but the clusters are not in equilibrium with each other.  Now 
the number of nuclei in each cluster is slowly changing and must be 
specified.  The abundances of species in cluster $j$ are then 
\begin{equation} 
Y^{(j)}(Z,A) = e^{\mu_h^{(j)}/kT} R_p^Z R_n^N Y^{NSE}(Z,A), 
\label{eq:qseabund2} 
\end{equation} 
where $\mu_h^{(j)}$ is the energy required to add a nucleus into cluster 
$j$ at constant entropy.   
 
As more of the nuclear reactions become slow due to the cooling, more QSE 
clusters appear.  For neutron-rich matter, the QSE clusters tend to break 
up into isotopic chains of a given $Z$, that is, clusters of nuclei 
in equilibrium 
under the exchange of neutrons but not protons or alpha particles.  This is 
the so-called $(n,\gamma)-(\gamma,n)$ equilibrium of the classical 
r-process phase.  Charged-particle strong and electromagnetic reactions have 
become too slow, so nuclei can only move from one $Z$ to the next now by 
either nuclear beta decay or neutrino capture.  The abundances in an 
$(n,\gamma)-(\gamma,n)$ equilibrium isotopic chain are simply found from 
Eq. (\ref{eq:qseabund2}) since these nuclei all belong to the same QSE 
cluster: 
\begin{equation} 
{Y(Z,A+1) \over Y(Z,A)} = \left({Y_n \over Y_{Qn}}\right)\left({m(Z,A+1) 
\over m(Z,A)}\right)^{3/2} \exp({S_n(Z,A+1)/kT}),       \label{eq:ngabund} 
\end{equation} 
where the neutron separation energy $S_n(Z,A+1)$ is given by 
\begin{equation} 
S_n(Z,A+1) = m(Z,A)c^2 + m_nc^2 - m(Z,A+1)c^2.    \label{eq:sndef} 
\end{equation} 
Eq. (\ref{eq:ngabund}) is the classic equation relating the abundances of 
neighboring isotopes in $(n,\gamma)-(\gamma,n)$ equilibrium \cite{B2FH}. 
 
The $(n,\gamma)-(\gamma,n)$ equilibrium eventually breaks down as 
neutron-capture and disintegration rates become too slow.  Even smaller QSE 
clusters appear (typically as adjacent pairs of even and odd $N$ isotopes), 
but these equilibria quickly break and the r-process freezes out.  The 
neutron-rich nuclei simply beta decay back to the stability line. 
 
\begin{figure} 
\centerline{\psfig{figure=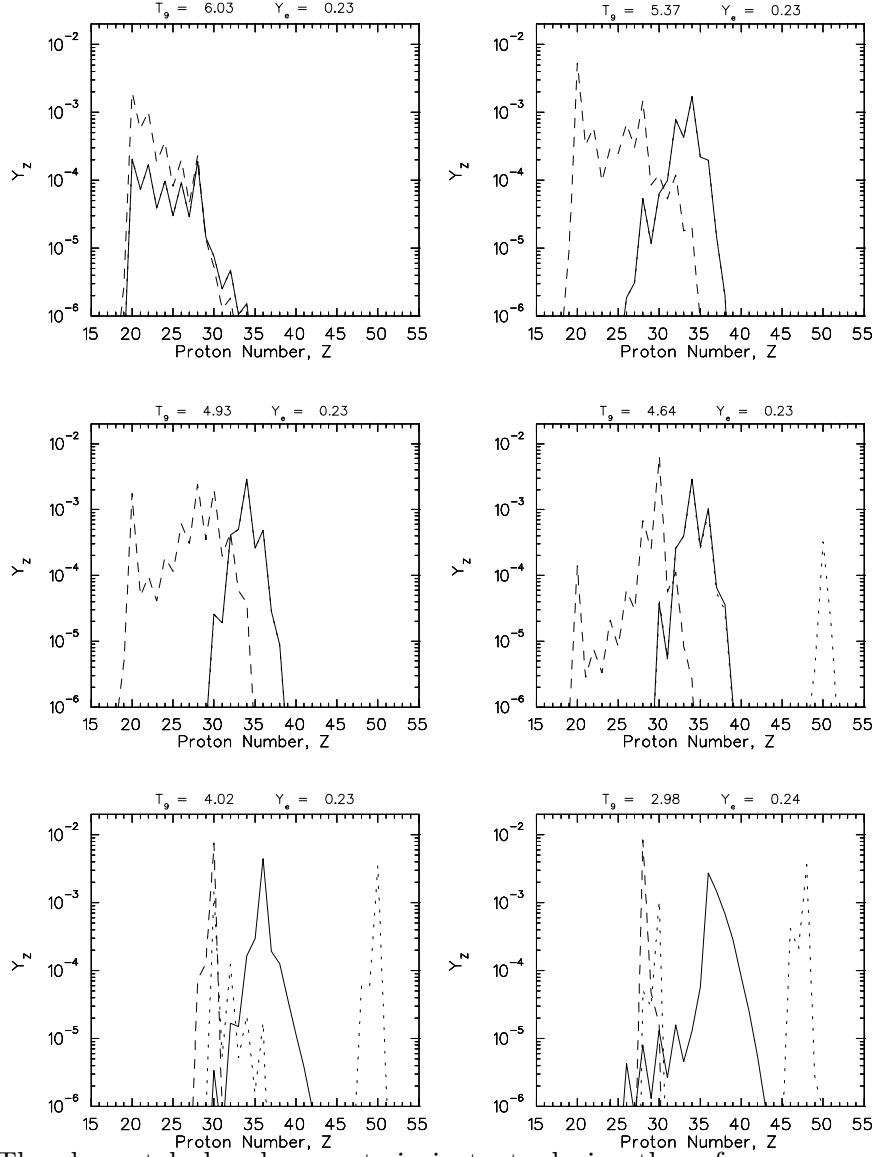,height=6.0in,angle=0}} 
\caption{ 
The elemental abundances at six instants during the reference expansion. 
In each panel, the solid curve gives the abundances from integration of the 
nuclear reaction network.  The dashed curve shows the NSE abundances for the 
same temperature, density, and $Y_e$ as in the network calculation.  The dotted 
curve shows the QSE abundances for the same temperature, density, $Y_e$, and 
$Y_h$ as in the network calculation.  By $T_9 =6.03$, the abundances have 
already fallen out of NSE.  However, they remain in QSE down below $T_9=5$. 
As the QSE distribution shifts to higher charge, the actual abundances cannot 
keep pace, and the system falls further in the hierarchy of statistical 
equilibria.  More QSE clusters develop.  The system eventually breaks down 
into $(n,\gamma)-(\gamma,n)$ equilibrium in which nuclei are in equilibrium 
only under exchange of neutrons.  This is the ``classical'' r-process phase 
of the expansion. 
} 
\label{fig:yz_model0} 
\end{figure}

Fig. \ref{fig:yz_model0} illustrates some of these ideas.  Shown are the 
elemental abundances for several temperatures during the reference expansion. 
The solid curve gives the abundances from the actual network calculations. 
The dashed curve shows the NSE abundances at the same temperature, density, 
and $Y_e$ as in the network calculation, while the dotted curve gives the QSE 
abundances at the same temperature, density, $Y_e$, and $Y_h$ as in the 
network calculation.  By $T_9 = 6.03$, the nuclear populations have already 
fallen out of NSE because of the slowness of the three-body reactions
assembling 
heavy nuclei.  The abundances are, however, very accurately in QSE.  The QSE 
is maintained through $T_9=4.93$, although the QSE and NSE are strongly 
diverging.  By $T_9 = 4.64$, an abundance peak at $Z=50$ is building up, but 
the network abundances are not keeping pace because of the slowness of the 
nuclear reactions that carry nuclei to higher charge.  The nuclei have fallen 
out of the large QSE cluster that contained all of the heavy nuclei. 
A larger number of more restricted QSE clusters is now present, so the system 
has dropped in the hierarchy of statistical equilibria.  By $T_9=4.02$, the 
abundance distribution is very different from that of the single large QSE. 
Interestingly, the abundances are dominated by the single isotope $^{94}$Kr 
with nearly 19\% of the mass (the remaining mass is in free neutrons and alpha 
particles).  This isotope serves as the initial seed nucleus for neutron 
captures during the subsequent ``classical'' r-process phase of the expansion. 
By $T_9 = 2.98$, some beta decays have already shifted matter to higher 
charge, but the r-process has only just begun.  An mpeg movie of Fig. 
\ref{fig:yz_model0} is available for viewing at the web site 
http://photon.phys.clemson.edu/movies.html.  Other movies at that site show 
the development of QSE clusters and the evolution of the abundances. 
 
\section{Nucleosynthesis Results} 
\label{sec:results}
 
We ran a total of eight models.  These are summarized in Table I.  In all 
cases the initial conditions were those of the reference calculation (model 
0) as were the neutrino temperatures and luminosities. In all cases,
at $T_9 = 10$ the material begins in weak steady state, which is set primarily 
by the neutrino and antineutrino capture reactions.
 Below $T_9=10$, the neutrino effects varied with the model.  For 
example, in 
model 1, neutrino capture on free nucleons and nuclei 
was disabled below $T_9=10$, but 
neutrino capture on heavy nuclei was turned back on for $T_9 \leq 3$.  For 
model 2, neutrino capture on both free nucleons and heavy nuclei was 
disabled below $T_9 = 10$, but both were turned back on for $T_9 \leq 3$. 
For model 3, neutrino capture occurred on free nucleons and heavy nuclei 
throughout the expansion. 
 
\begin{figure} 
\centerline{\psfig{figure=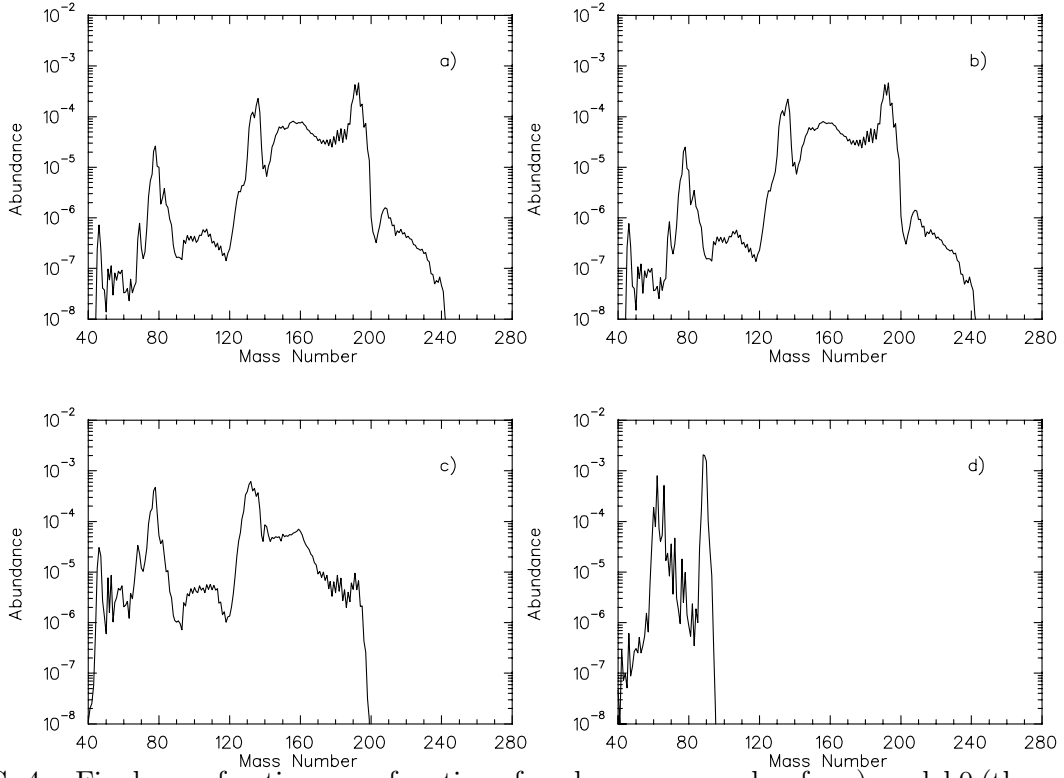,height=4.0in,angle=90}} 
\caption{ 
Final mass fractions as a function of nuclear mass number for a) model 0 (the 
reference calculation), b) model 1 (neutrino and anti-neutrino capture only on 
heavy nuclei for $T_9 \leq 3$), c) model 2 (neutrino and anti-neutrino 
capture on free nucleons and heavy nuclei for $T_9 \leq 3$), and d) model 3 
(neutrino and anti-neutrino capture on free nucleons and nuclei 
throughout the expansion).  A robust r-process has occurred for 
models 0 and 1 allowing production of the $A=195$ peak nuclei.  In model 2, the 
resulting r-process is less robust--the $A=195$ is not present.
In model 3 there is no production of heavy r-process nuclei. 
} 
\label{fig:abund} 
\end{figure}

We investigated in detail models 0 through 3 to see the effects of neutrino 
capture during the r-process.  Fig. \ref{fig:abund} shows the final
abundances versus nuclear mass number for each of these models.
There is only a slight difference between models 0 (no neutrino effects 
below $T_9=10$) and 1 (neutrino capture on heavy nuclei below $T_9=3$). 
The curves show strong $A=130$ and $A=195$ abundance peaks, with the latter
larger in abundance.  These models have experienced a robust r-process.
By contrast, model 2 (neutrino capture on heavy nuclei and 
free nucleons below $T_9=3$) shows an $A=130$ abundance peak, but neutrino 
captures on free nucleons during the r-process phase has prevented the run 
up to $A=195$.  Finally, model 3 (neutrino capture on heavy nuclei and 
free nucleons on throughout the expansion) shows no r-process.  The mass is 
largely concentrated in three nuclear species $^{88}$Sr, $^{89}$Y, and
$^{90}$Zr.  Neutrino capture has completely prevented the r-process in
this case. 
 
\begin{figure} 
\centerline{\psfig{figure=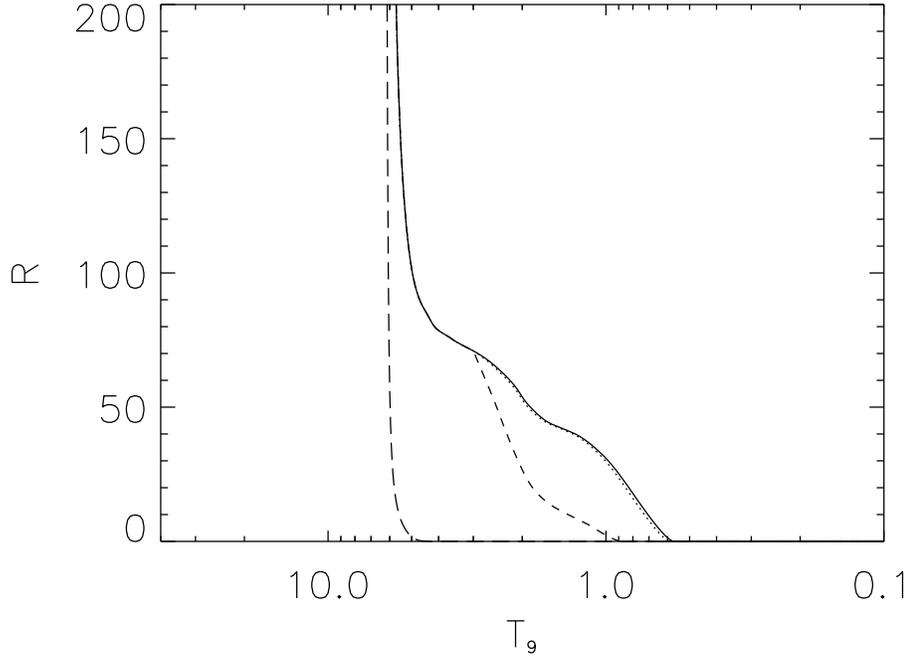,height=4.0in,angle=90}} 
\caption{ 
The neutron-to-seed ratio $R$ in models 0 (solid curve), 1 (dotted curve), 
2 (short-dashed curve), and 3 (long-dashed curve).  Neutrino capture on 
free nucleons strongly reduces $R$ during the expansion and thereby hampers 
the r-process. 
} 
\label{fig:R} 
\end{figure}

The ``success'' of an r-process expansion 
in making heavy species depends on $R$, the 
ratio of the abundance of free neutrons to that of the heavy seed nuclei that 
capture those free neutrons during the r-process.  The larger $R$ is, the 
more neutrons each nucleus will capture on average, and, consequently, the 
heavier the final nuclei produced.  Fig. \ref{fig:R} shows $R$ in models 0 
through 3.  In models 0, 1, and 2, $R$ is about 70 at $T_9=3$, roughly the 
beginning of the r-process phase of the expansion.  Since the average heavy 
nucleus has mass number $\sim$100 at this point, the final average heavy
nucleus has a mass number $\sim 170$.  In the reference 
calculation (model 0), $R$ declines gradually as the temperature falls. 
This is due to the capture of neutrons by nuclei during the r-process. 
Model 1 follows suit, although the neutrino capture by heavy nuclei 
for $T_9 = 3$ 
slightly enhances the rate of increase of nuclear charge and thereby lowers 
$R$ a little for each temperature.  In model 2, however, $R$ plummets 
drastically once the neutrino captures are enabled below $T_9 = 3$.  The 
only difference between models 1 and 2 is that model 2 includes neutrino 
capture on free nucleons which must therefore be the cause of the large 
drop in $R$.  Finally, in model 3, $R$ drops to zero before the r-process 
can even begin. 
 
\begin{figure} 
\centerline{\psfig{figure=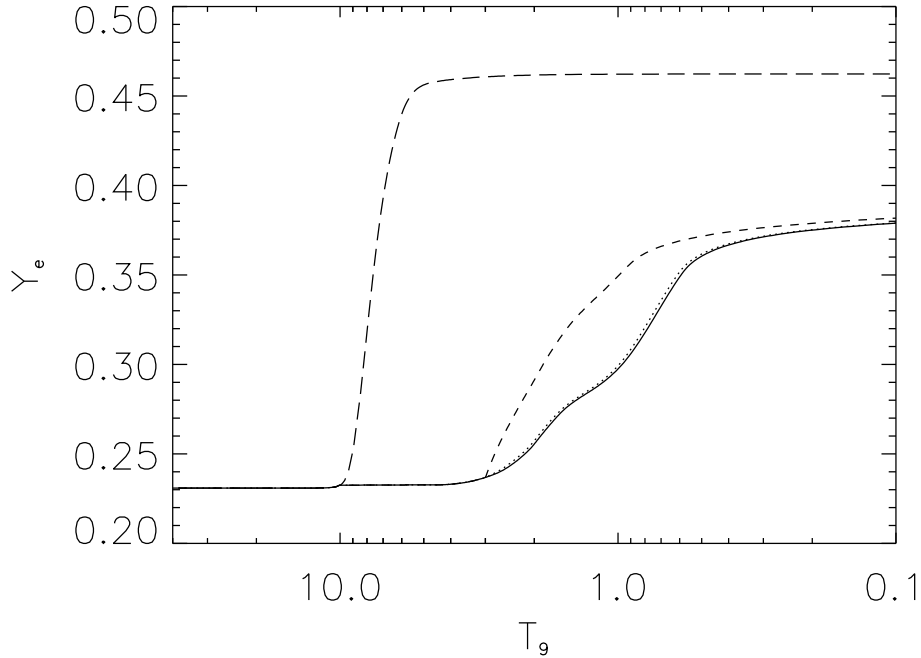,height=4.0in,angle=90}} 
\caption{ 
The electron fraction $Y_e$ for models 0 (solid curve), 1 (dotted 
curve), 2 (short-dashed curve), and 3 (long-dashed curve).  In the 
reference expansion (model 0), $Y_e$ starts to grow as the r-process 
phase of the nucleosynthesis begins for $T_9 < 3$.  It rises due to nuclear 
beta decays.  From model 1, it is apparent that neutrino captures on heavy 
nuclei slightly enhance $Y_e$ during the expansion.  Model 2, however, shows 
that neutrino capture on free neutrons (for $T<3$) has a much larger effect 
on $Y_e$ than capture on heavy nuclei.  Model 3 shows that neutrino capture 
on free neutrons early has an even greater influence due to the ``alpha effect''. 
} 
\label{fig:ye} 
\end{figure}

The effects of neutrino capture are also apparent in Fig. \ref{fig:ye}, 
which shows $Y_e$ for models 0 through 3.  $Y_e$ is set early in the expansion 
by the interactions of neutrinos and anti-neutrinos with free neutrons and 
protons.  The anti-neutrinos, which capture on protons to make neutrons, 
have a higher temperature and luminosity
than the neutrinos, which convert neutrons 
into protons.  This gives rise to the low steady-state value of $Y_e$. 
For model 0, $Y_e$ stays low until the r-process begins at about $T_9 = 3$, 
then $Y_e$ rises due to the nuclear beta decays that increase the nuclear 
charge.  Model 1 shows similar behavior, although $Y_e$ increases slightly 
faster due to the added effect of neutrino capture on heavy nuclei.  By 
contrast, model 2 shows a much more rapid rise in $Y_e$ below $T_9 = 3$. 
Again, this is due to neutrino capture on free neutrons.  We note that 
the number of free neutrons per nucleon at $T_9=3$ is 0.38 while the number 
of neutrons locked up into heavy nuclei is 0.34.  A neutrino capture by
either a bound or free neutron increases $Y_e$ by the same amount, and the
number of bound and free neutrons is about the same.
The rather strong difference between models 1 and 2 is due
to the much larger cross section for neutrino capture on free neutrons than 
on bound neutrons due to the larger number of final states available to 
the former.  Neutrino capture on a free neutron decreases the 
number of neutrons available for incorporation into nuclei.  For these 
simple reasons, neutrino capture during the r-process on balance must hinder 
the production of high-mass nuclei.  Model 3 shows an even more dramatic 
rise in $Y_e$.  For this model, $Y_e$ quickly rises to about 0.45 and 
changes little for the rest of the expansion.  This drastic behavior
partly reflects the classic alpha effect identified in 
\cite{fullermeyer95,mfw96}. 
 
\begin{figure} 
\centerline{\psfig{figure=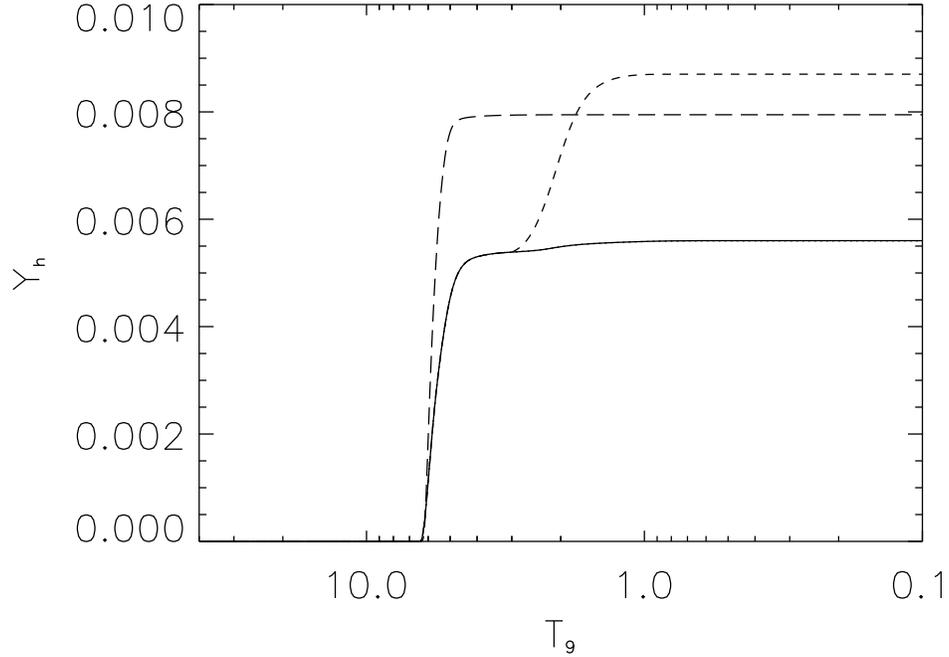,height=4.0in,angle=90}} 
\caption{ 
The abundance of heavy nuclei (i.e. those nuclei with $A\geq12$) per 
nucleon for models 0 (solid curve), 1 (dotted curve), 2 (short-dashed 
curve), and 3 (long-dashed curve).  Neutrino capture on free nucleons 
enhances production of heavy seed nuclei.  This strongly reduces the 
neutron-to-seed ratio $R$ and limits the r-process.  Notice the rise in 
$Y_h$ in the reference calculation (model 0) for $T_9 \leq 2$.  This 
results from the late assembly of alpha particles into heavy nuclei via the 
$\alpha+\alpha+n \to ^9$Be followed by $^9{\rm Be}(\alpha,n)^{12}$C 
reaction sequence.
} 
\label{fig:yh} 
\end{figure}

Neutrino capture on free neutrons does not simply hinder the possibility of 
making high-mass nuclei by depleting the supply of free neutrons.  It also 
increases $Y_h$, the number of heavy nuclei.  This in turn lowers $R$. 
Fig. \ref{fig:yh} shows $Y_h$ in models 0 through 3.  In the reference 
calculation (model 0), $Y_h$ rises to about 0.005 at $T_9 = 5$ and then 
increases slowly from that point on.  The small rise for $T_9 \approx 3$ 
comes from late assembly of heavies via the $\alpha + \alpha +n$ reaction.
Because $Y_h$ changes only slowly 
for $T_9 < 5$, the nuclear populations are in a large QSE.  The subsequent 
evolution is to break that large QSE into smaller QSE clusters and 
eventually $(n,\gamma)-(\gamma,n)$ equilibrium, as described in Sec.
\ref{sec:hier}.  

Model 1 behaves exactly the same as model 0, and there is
no perceptible difference in the number of heavy nuclei in the two models.
In model 2, however, the effect of neutrino capture on free 
neutrons dramatically increases the number of heavy nuclei for $T_9 < 3$.  The 
exact mechanism for production of new heavy nuclei is discussed below
in Sec. \ref{sec:heavy}.  In 
model 3, the number of heavy nuclei shoots up between $T_9 = 6$ and $T_9=5$ 
to a large value and then stays constant. 
 
\begin{figure} 
\centerline{\psfig{figure=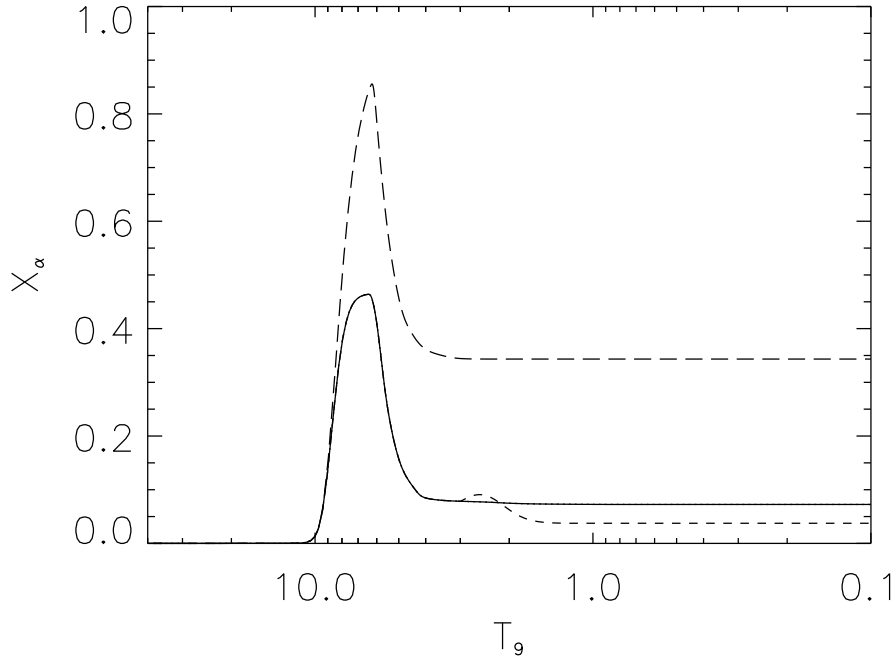,height=4.0in,angle=90}} 
\caption{ 
The mass fraction of alpha particles in models 0 (solid curve), 1 (dotted 
curve), 2 (short-dashed curve), and 3 (long-dashed curve).  In the 
reference calculation (model 0), the alpha mass fraction grows to a value 
of about 0.45 ($T_9\approx 7$) as neutrons and protons assemble into alpha 
particles and then falls again as the alphas assemble into heavier nuclei. 
Models 1 and 2 follow suit, although neutrino capture on free neutrons 
affects the alpha mass for $T_9 < 3$ in model 2.  For model 3, the alpha 
mass fraction grows to a value larger than 0.8.  This is the ``alpha 
effect''.  As protons lock up into $^4$He nuclei, which are essentially 
inert to neutrino capture, ongoing neutrino capture on free neutrons leads 
to production of new protons which in turn make new alphas.  This strongly 
enhances the resulting alpha mass, increases $Y_e$, and reduces $R$, the 
neutron-to-seed ratio. 
} 
\label{fig:xa} 
\end{figure}

The reason for the sudden rise in $Y_h$ for model 3 may be seen in Fig. 
\ref{fig:xa}, which presents the mass fraction of alpha particles for 
models 0 through 3.  First we discuss the other models. 
The evolution of the alpha particle mass fraction, 
$X_\alpha$, in models 0, 1, and 2 is 
nearly the same.  As $T_9$ drops below 10, neutrons and protons begin to 
assemble into alpha particles.  $X_\alpha$ peaks at 0.46 at about $T_9=7$. 
The remaining mass is in free neutrons.  This correctly gives $Y_e=X_\alpha/2 
=0.23$ (cf. Fig. \ref{fig:ye}). $X_\alpha$ then 
falls as the alphas assemble into heavy nuclei.  This assembly of heavy 
nuclei slows down considerably below $T_9 = 4$, and the alphas freeze out 
with a final mass fraction of about 0.07.  In model 2, the neutrino 
captures on free neutrons causes the alpha mass fraction at first to rise 
and then to fall.  This is related to the production of new heavy nuclei 
seen in Fig. \ref{fig:yh}.  In model 3, however, $X_\alpha$ rises to a much 
higher value and also freezes out at a much higher level.  The reason for 
this is the so-called ``alpha effect'' \cite{fullermeyer95}. 
 
The alpha effect occurs when neutrons and protons assemble into $^4$He in 
the presence of a large flux of neutrinos.  The interaction of 
neutrinos with neutrons and anti-neutrinos with protons sets up a 
steady-state ratio of the abundance of free neutrons to protons $Y_n/Y_p$. 
In the present models we have initially $Y_n/Y_p=3.3$. 
At high temperature, only free nucleons are present, so $Y_n/Y_p = 3.3$ 
corresponds to $Y_e = 0.23$.  However, as protons lock up into 
$^4$He, which is largely inert to the neutrino interactions, 
$Y_n/Y_p$ rises rapidly. 
Neutrino 
captures on free neutrons will tend to reset $Y_n/Y_p$ back to 3.3, in 
accordance with the neutrino temperatures.  The newly created protons, 
however, do not remain free but rather gather into new alpha particles, 
again upsetting the $Y_n/Y_p$ ratio.  This then establishes a runaway that 
causes the dramatic rise in $X_\alpha$ and $Y_e$.  $Y_e$ would rise to 0.5 
were it not for the fact that heavy nuclei form and soak up free neutrons, 
thereby limiting the alpha effect runaway.  In any event, the alpha effect 
causes a rapid depletion of free neutrons that kills the possibility of an 
r-process. 
 
\begin{figure} 
\centerline{\psfig{figure=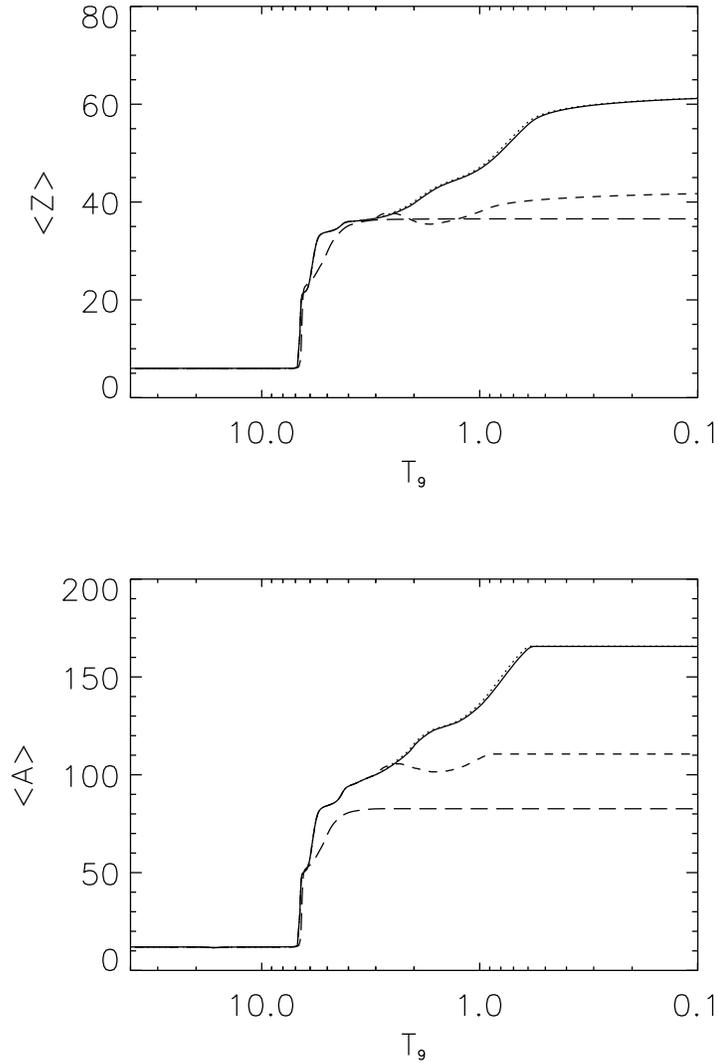,height=6.0in,angle=0}} 
\caption{ 
The average charge $<Z>$ (upper panel) and average mass number $<A>$ 
(lower panel) of heavy nuclei in models 0 (solid curve), 1 (dotted 
curve), 2 (short-dashed curve), and 3 (long-dashed curve).  In models 0, 1, 
and 2, the seed nuclei build up to about charge 40 and mass 100 by the 
onset of the r-process at $T_9 \approx 3$.  In models 0 and 1, nuclear beta 
decays and neutron captures during the r-process then increase $<Z>$ and 
$<A>$.  In model 2, however, neutrino capture on free nucleons creates new 
and lighter seed nuclei.  This causes the average charge and mass to drop 
before climbing again as the assembly of new nuclei ceases.  In model 3, 
the average charge and mass freeze out early because the alpha effect has 
depleted the supply of light particles. 
} 
\label{fig:za1} 
\end{figure}

The average charge $ \langle Z \rangle$ and mass number 
$\langle A \rangle$ of heavy nuclei in each of 
the four models are shown in Fig. \ref{fig:za1}.  The initial build up of 
heavy nuclei between $T_9 = 7 $ and $T_9 = 5$ is apparent, as is the 
subsequent r-process for models 0 and 1.  There is no discernible
difference between the final average charge and mass in these two models
although neutrino captures do cause the nuclei to work their way up the
network faster in model 1.  Model 2 has an interesting dip in 
$ \langle Z \rangle$ 
and $ \langle A \rangle$ for $T_9$ between 3 and about 0.5.  
This results from the 
assembly of new heavy nuclei already seen in Fig. \ref{fig:yh}.  This 
creates new and much lighter nuclei than were present for $T_9 > 3$, 
thereby lowering the average charge and mass.  Only as the assembly of new 
seed nuclei shuts off below $T_9 = 2$ can these quantities again rise by 
r-processing.  The damage has already been done, however, and the final average 
charge and mass in model 2 are much less than in models 0 and 1.  
$ \langle Z \rangle$ and 
$ \langle A \rangle$ freeze out in model 3 at about $T_9 = 5$.  
In this case, the alpha 
effect depletes the abundance of light particles and there is little 
subsequent evolution in the abundances. 
 
In summary, neutrino capture during nucleosynthesis severely hinders the 
r-process in the present models.
When all neutrino capture effects during nucleosynthesis 
are included (model 3), no 
r-process occurs at all, even though the same calculation without neutrino 
capture (model 0) 
at all yields an extremely robust r-process.  The tiny helpful effects 
of capture on heavy nuclei (model 1) 
are more than offset by the detrimental effects 
of capture on free neutrons. 
This is true even if neutrino capture on free neutrons only 
occurs during the r-process phase (model 2). 
 
\section{Assembly of Heavy Nuclei}   \label{sec:heavy}
 
The results in the previous section showed that neutrino-capture enhanced 
assembly of heavy nuclei strongly hindered the r-process.  In this section 
we study this in more detail.  The goal is to understand exactly how the 
neutrino capture actually hinders the r-process. 
 
In the alpha effect, neutrino capture drastically reduces the abundance of 
free neutrons.  It also increases the abundance of alpha particles, which 
in turn allows assembly of more seed nuclei.  Which of the two effects, 
loss of neutrons or assembly of new seed nuclei, is the dominant one in 
limiting the r-process? 
 
We can understand this as follows.  The rate 
of change of the neutron-to-seed ratio $R$ is
\begin{equation} 
{dR \over dt} = {d(Y_n/Y_h) \over dt} = {1 \over Y_h} {dY_n \over dt} 
-{Y_n \over Y_h^2}{dY_h \over dt}.  \label{eq:dRdt} 
\end{equation} 
In order to compute $dY_n/dt$ and $dY_h/dt$, we must consider the fate of a 
proton newly formed by a neutrino capture on a free neutron.  Such a proton 
will most likely capture two neutrons to become a tritium nucleus.  The 
tritium may then capture another neutrino-capture produced proton in a 
$t + p \to n + ^3$He reaction.  The $^3$He will then quickly capture 
another neutron to become $^4$He.  Alternatively, the tritium may capture 
another tritium in the reaction $t + t \to 2n + ^4$He.  In either case, each  
neutrino capture on a free neutron leads to the disappearance of two free 
neutrons--the neutron suffering the neutrino capture and the neutron that 
accompanies it into $^4$He.  For this reason, 
\begin{equation} 
{dY_n \over dt} = - 2 \lambda Y_n,  \label{eq:dyndt} 
\end{equation} 
where $\lambda$ is the rate of neutrino capture on free neutrons.
In principle we should include creation of neutrons by antineutrino capture
on protons.  However the mass fraction of protons
becomes negligible as soon as alpha particles form, so we neglect this
process in Eq. (\ref{eq:dyndt}).  Furthermore, we neglect in Eq.
(\ref{eq:dyndt}) the loss of free neutrons due to incorporation into nuclei
during the r-process.  This loss occurs slowly if the neutrino capture is
large.
  
$^4$He nuclei assembled in this way will tend to react via a three-body 
reactions to form new heavy nuclei.  
At late time, $T_9 < 6$, some alpha
particles will be in existence while some will be created after neutrino
captures on neutrons.  If $n$ (where on average $n \lesssim 6$)  
neutrino captures 
are required to produce a new heavy nucleus then, 
\begin{equation} 
{dY_h \over dt} = {\lambda Y_n \over n}.  \label{eq:dyhdt} 
\end{equation} 
Substitution of Eqs. (\ref{eq:dyndt}) and (\ref{eq:dyhdt}) into Eq. 
(\ref{eq:dRdt}) then yields 
\begin{equation} 
{dR \over dt} = -2 \lambda R - { \lambda R^2 \over n}.  \label{eq:dR2dt} 
\end{equation} 
 
The $R^2$ term in Eq. (\ref{eq:dR2dt}) arises from the assembly of new seed 
nuclei.  It provides the dominant reduction in $R$ throughout most of the 
expansion.  For example, early in the expansion, when $R=300$, say, the 
assembly of new seed nuclei induced by neutrino capture causes $R$ to 
decline at a rate about 15,000 times faster than the rate of
neutrino capture on 
free neutrons.  This explains the extremely fast drop in $R$ in model 3 
seen in Fig. \ref{fig:R}. 
Only as $R$ drops below 10 does the simple disappearance of 
neutrons cause a larger drop in $R$.  These considerations explain why the 
alpha effect is so devastating to the r-process. 
 
Model 2 also shows a rapid drop in $R$ and a rapid 
increase in the number of heavy nuclei after neutrino capture on free 
neutrons turns on at $T_9=3$.  This is perhaps surprising since the 
three-body reactions assembling heavy nuclei from alpha particles are 
rather slow at this temperature.  It indicates that 
another channel has opened 
for the assembly of seed nuclei. 
 
Another possible fate of a tritium nucleus formed via neutrino capture on 
a free neutron is to capture one of the 
abundant alpha particles to become $^7$Li which can then capture other 
light particles to assemble new heavy nuclei.  This does not happen 
immediately in model 2, however.  At $T_9\approx 3$, the $(\gamma,\alpha)$ 
reaction on $^7$Li is rapid and keeps the net $^3{\rm 
H}(\alpha,\gamma)^7{\rm Li}$ rate low.  This allows tritium to move into 
$^4$He.  As the temperature drops, the disintegration of $^7$Li slows, and 
tritium increasingly captures to $^7$Li.  This leads 
to significant assembly of 
new heavy nuclei (Fig. \ref{fig:yh}), which again poisons the 
r-process by decreasing $R$. 
 
\begin{figure} 
\centerline{\psfig{figure=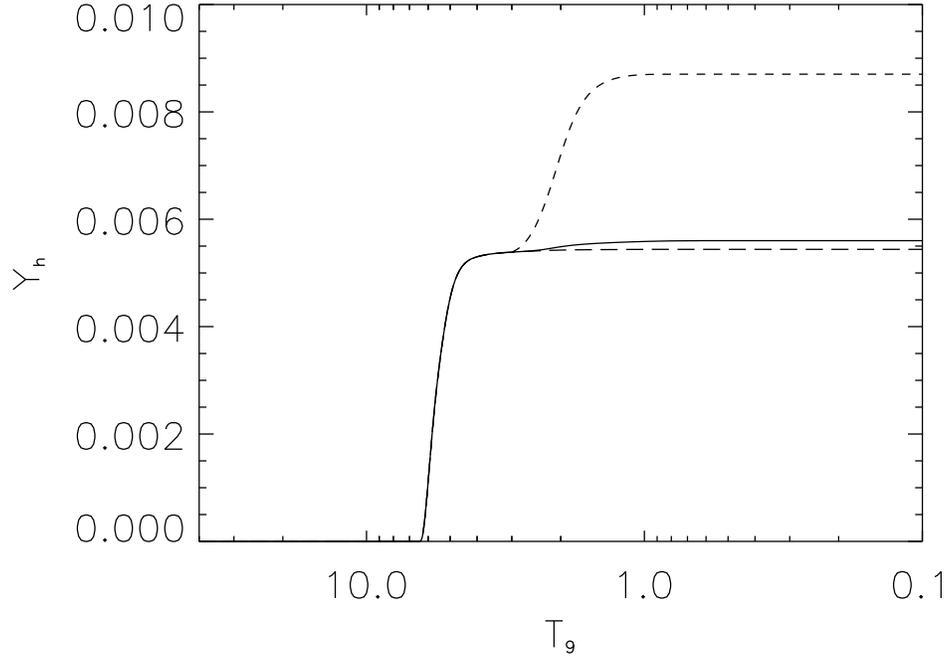,height=4.0in,angle=90}} 
\caption{ 
The abundance of heavy nuclei $Y_h$ in models 0 (solid line), 2 
(short-dashed curve), and 4 (long-dashed curve).  Model 4 is identical to 
model 2 except that the reaction $^3{\rm H}(\alpha,\gamma)^7{\rm Li}$ has 
been disabled.  This prevents assembly of new seed nuclei induced by 
neutrino capture on free neutrons.  When the $^7$Li channel is open (model 2), 
$Y_h$ grows dramatically. 
} 
\label{fig:yhnoli} 
\end{figure}

\begin{figure} 
\centerline{\psfig{figure=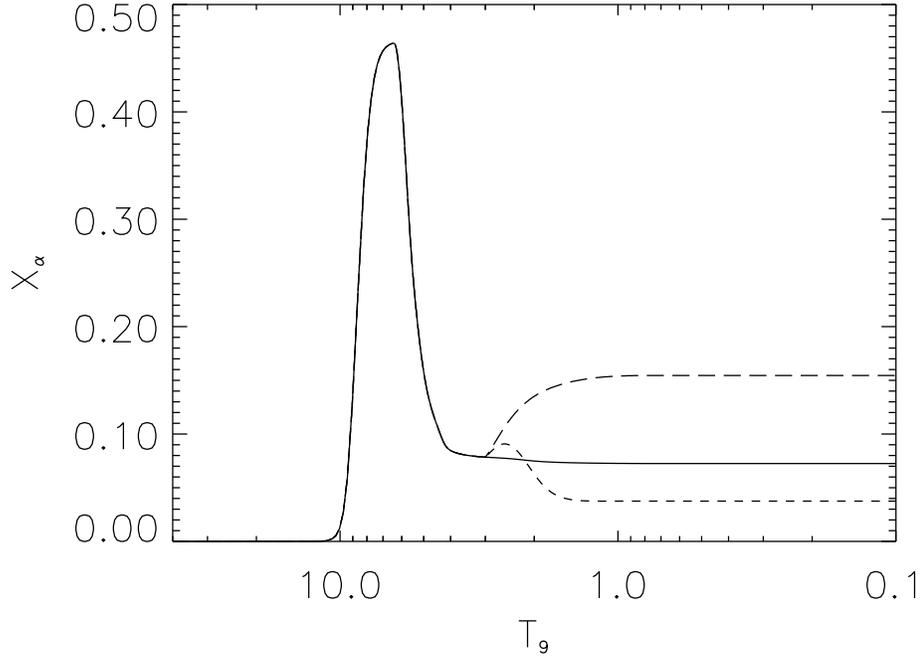,height=4.0in,angle=90}} 
\caption{ 
The mass fraction of $^4$He in models 0 (solid line), 2 
(short-dashed curve), and 4 (long-dashed curve).  Model 4 is identical to 
model 2 except that the reaction $^3{\rm H}(\alpha,\gamma)^7{\rm Li}$ has 
been disabled.  This prevents assembly of new seed nuclei induced by 
neutrino capture on free neutrons.  The fate of a proton produced 
from neutrino capture on a free neutron is thus to end up as a $^4$He 
nucleus, hence the rise in $X_\alpha$ in model 4. 
} 
\label{fig:xanoli} 
\end{figure}

That this is indeed the mechanism for assembly of new heavy nuclei is 
apparent from Fig. \ref{fig:yhnoli}.  This figure shows $Y_h$ during models 
0 (solid curve), 2 (short-dashed curve), and 4 (long-dashed curve).  Model 
4 was identical to model 2 (neutrino capture on free nucleons and heavy 
nuclei below $T_9=3$) except that the $t+^4{\rm He} \to ^7$Li reaction was 
disabled.  Few new seed nuclei assemble in model 4 because the three-body 
reactions are slow and the $^7$Li channel is closed.  
In this case, neutrino capture on 
free neutrons simply leads to a substantial increase in the mass fraction 
of alpha particles, as is apparent in Fig. \ref{fig:xanoli}.  Because the 
free neutrons rapidly disappear in model 4 below $T_9=3$, the 
$\alpha+\alpha+n$ reaction is very slow, and model 4 ends up with even 
fewer heavy nuclei than model 0.  In spite of this, the neutrino 
capture means the r-process is less robust in model 4 than in model 0. 
 
As a final point, the late assembly of new seed nuclei through the $^7$Li 
channel will happen even if no $^4$He nuclei initially are present.  Once 
neutrino capture on free neutrons turns on, the $^4$He abundance will grow 
until it is sufficiently large to allow the $^3{\rm H} 
(\alpha,\gamma)^7$Li reaction 
to proceed rapidly.  Thus, late assembly of seed nuclei will occur even in 
low-entropy r-processes (for which the alpha abundance is initially very low) 
if the neutrino flux is large. 
 
\section{Neutrino Capture and QSE}    \label{sec:qse} 
 
The alpha effect and ongoing neutrino capture on free neutrons essentially 
precludes the possibility of accelerating the r-process via 
strong neutrino capture on heavy nuclei.  Nevertheless, it is interesting 
to consider what the effect of neutrino capture on heavy nuclei alone 
would do in an intense neutrino flux. 
 
One might suppose that allowing neutrino capture on heavy nuclei 
to occur earlier in the expansion than $T_9=3$ as in model 1 would help the 
r-process.  Earlier neutrino capture would move nuclei to higher mass 
earlier and possibly accelerate the r-process.  In fact the opposite is true 
and points out the importance of the QSE concept in considerations of 
neutrino capture during the r-process. 
 
To test the effect of neutrino capture on heavy nuclei,
we ran models 5, 6, and 7, which were identical to model 1 except 
that they used a larger neutrino luminosity of $L_{\nu} = 10^{52}$ ergs/s 
for electron neutrinos and $L_{\bar{\nu}_e} = 4 \times 10^{52}$ ergs/s for
the electron anti-neutrinos
and that, respectively, they allowed neutrino capture only on heavy 
nuclei to begin 
at $T_9=7$, $T_9=5$, and $T_9=3$.  The larger luminosity increases the 
neutrino capture and thereby shows the effects more clearly.  (Indeed,
without the increased neutrino luminosities, we saw little effect even if
the neutrino capture on nuclei was on throughout the expansion.)  Since no
heavy nuclei exist prior to $T_9 = 7$, model 5 is equivalent to having 
neutrino
capture on heavy nuclei on throughout the expansion.
 
\begin{figure} 
\centerline{\psfig{figure=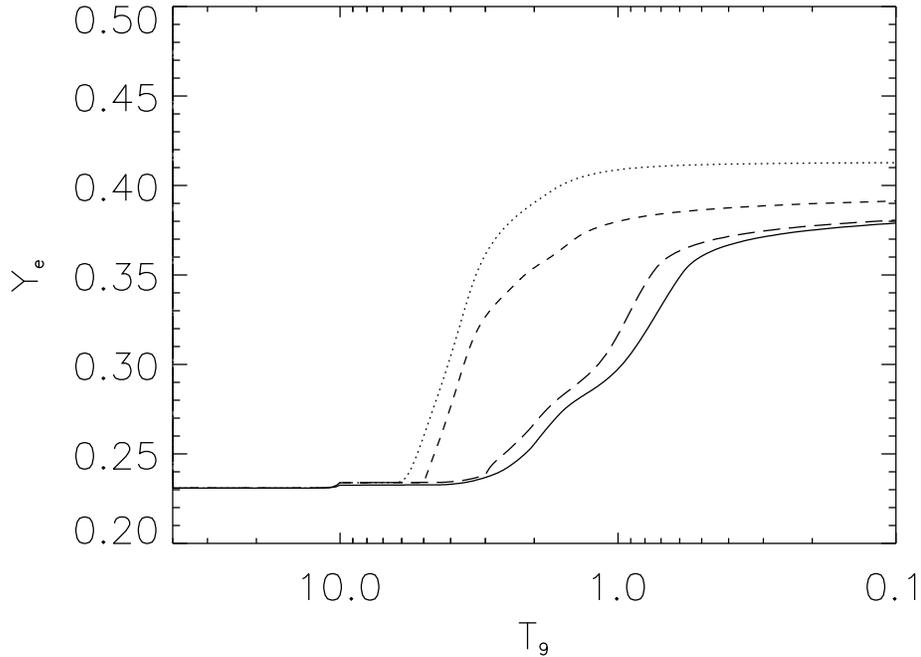,height=4.0in,angle=90}} 
\caption{ 
$Y_e$ during the expansion for models 0 (solid curve), 5 (dotted curve), 6 
(short-dashed curve), and 7 (long-dashed curve).  In models 5, 6, and 7, 
there is only neutrino capture on heavy nuclei for $T_9<7$, $T_9 < 5$, and 
$T_9 < 3$, respectively.  The earlier the neutrino capture is allowed to 
occur, the earlier $Y_e$ rises.  This does not translate into an increase
in the average charge and mass of the heavy nuclei, however (see 
Fig. 14). 
} 
\label{fig:yelarge} 
\end{figure}

\begin{figure} 
\centerline{\psfig{figure=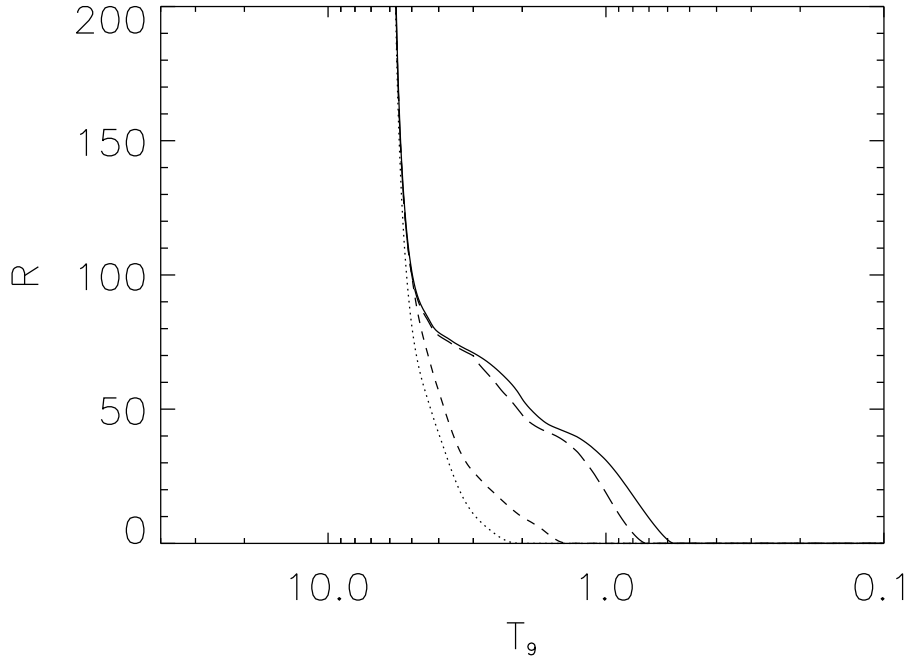,height=4.0in,angle=90}} 
\caption{ 
$R$ during the expansion for models 0 (solid curve), 5 (dotted curve), 6 
(short-dashed curve), and 7 (long-dashed curve).  In models 5, 6, and 7, 
there is only neutrino capture on heavy nuclei for $T_9<7$, $T_9 < 5$, and 
$T_9 < 3$, respectively.  $R$ declines sharply due to neutrino capture on 
heavy nuclei alone.  This does not lead to a more robust r-process, however. 
When the neutrino capture occurs during the QSE phase of the expansion,
the equilibrium among heavy nuclei allows neutrons 
to convert rapidly into protons without increasing the average charge.
This decreases $R$ and hinders the r-process. 
} 
\label{fig:Rlarge} 
\end{figure}

\begin{figure} 
\centerline{\psfig{figure=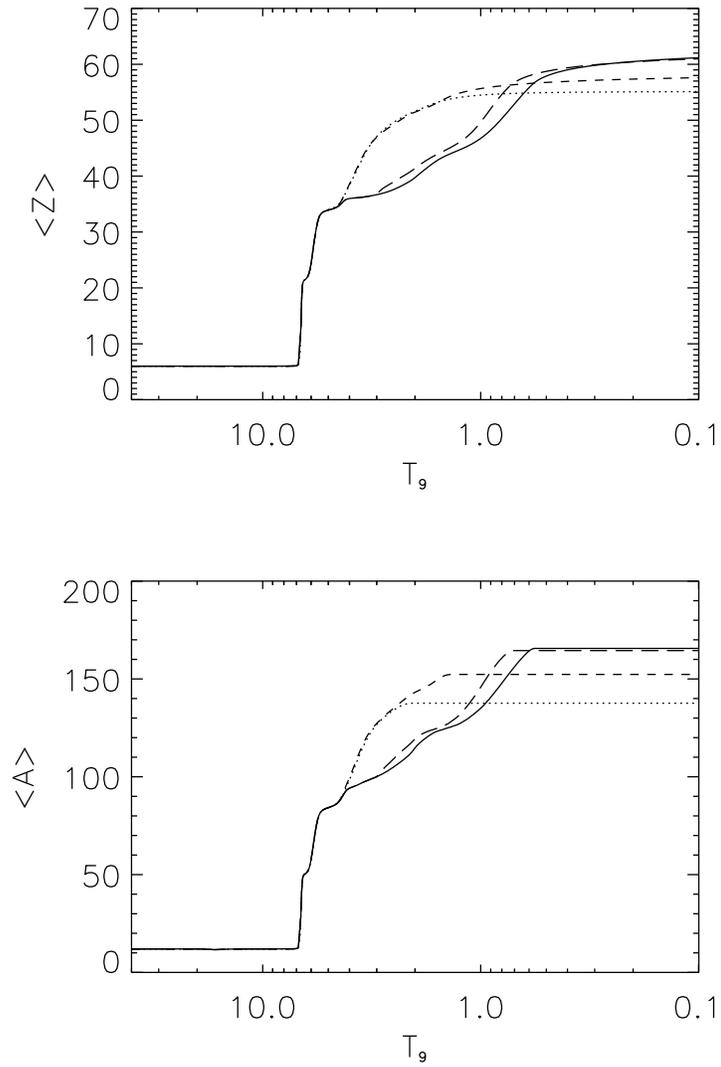,height=6.0in,angle=0}} 
\caption{ 
Average charge $<Z>$ (top panel) and average mass number $<A>$ (bottom 
panel) of heavy nuclei 
during the expansion for models 0 (solid curve), 5 (dotted curve), 6 
(short-dashed curve), and 7 (long-dashed curve).  Stronger neutrino 
capture by heavy nuclei limits the r-process if the captures happen during 
the QSE phases of the expansion.
} 
\label{fig:ZAlarge} 
\end{figure}

Fig. \ref{fig:yelarge} compares $Y_e$ in models 5, 6, and 7 
to that in model 0.  As is evident, allowing neutrino capture 
to occur earlier does cause $Y_e$ to rise much earlier in the expansion. 
It also causes $R$ to drop much more rapidly, as seen in Fig. \ref{fig:Rlarge}. 
This does not, however, correspond to a more robust r-process in models 
5 and 6.  Fig. \ref{fig:ZAlarge} shows $<Z>$ and $<A>$ for models 0, 5, 6, 
and 7, and it is readily apparent that early neutrino capture on heavy nuclei 
alone in fact tends to hinder the r-process.  This result may seem
counter-intuitive, so we study it in a little more detail. 
 
As discussed in Sec. \ref{sec:hier},
the nuclei begin in NSE at high temperature. 
As the temperature falls, the NSE breaks down (at $T_9\approx 6-7$ in the 
present calculations).  The system descends the hierarchy of statistical 
equilibria and goes into a QSE in which the heavy nuclei are all in equilibrium 
under exchange of light particles.  In the present calculations, a large QSE 
containing the most abundant nuclei
persists down to $T_9 \approx 4.0$ after which point an increasing number
of smaller QSE clusters appears.  Eventually these evolve to the 
$(n,\gamma)-(\gamma,n)$ equilibria that represent the classical r-process 
phase of the expansion.  The crucial point for neutrino capture is that even 
below $T_9 \approx 5$, the nuclear abundances are interlocked.  This means 
that any increase in $Y_e$ via a neutrino capture can cause a readjustment of 
all nuclear abundances, not just that of the neutrino-capturing nucleus. 
 
\begin{figure} 
\centerline{\psfig{figure=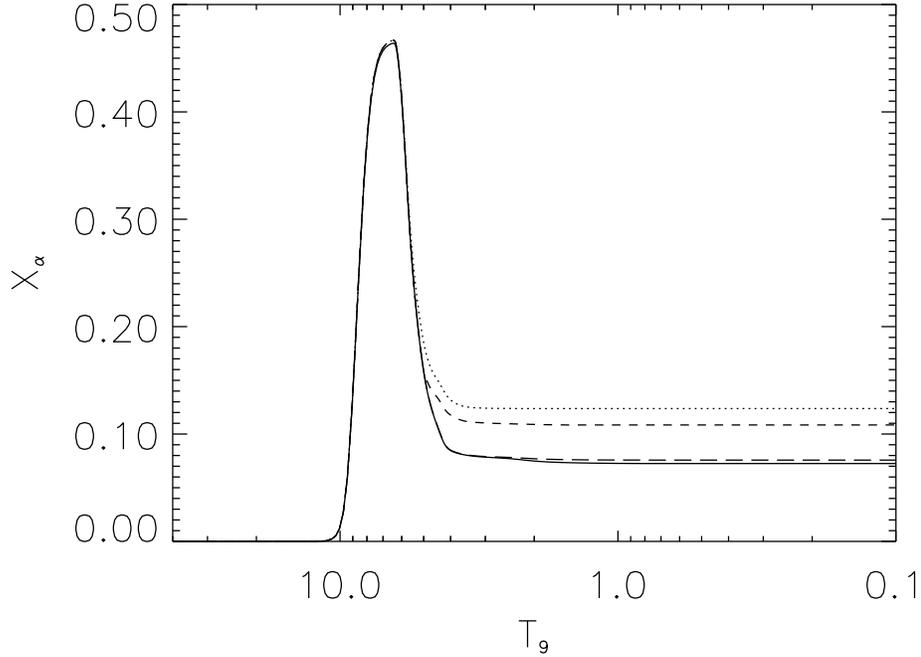,height=4.0in,angle=90}} 
\caption{ 
The mass fraction of alpha 
particles during the expansion for models 0 (solid curve), 5 (dotted curve), 6 
(short-dashed curve), and 7 (long-dashed curve).  In models 5, 6, and 7, 
there is only neutrino capture on heavy nuclei for $T_9<7$, $T_9 < 5$, and 
$T_9 < 3$, respectively.  Neutrino captures on heavy nuclei increases 
$Y_e$.  This tends to increase the abundance of free protons in the QSE. 
The protons capture neutrons and increase the abundance of $^4$He. 
} 
\label{fig:xalarge} 
\end{figure}

\begin{figure} 
\centerline{\psfig{figure=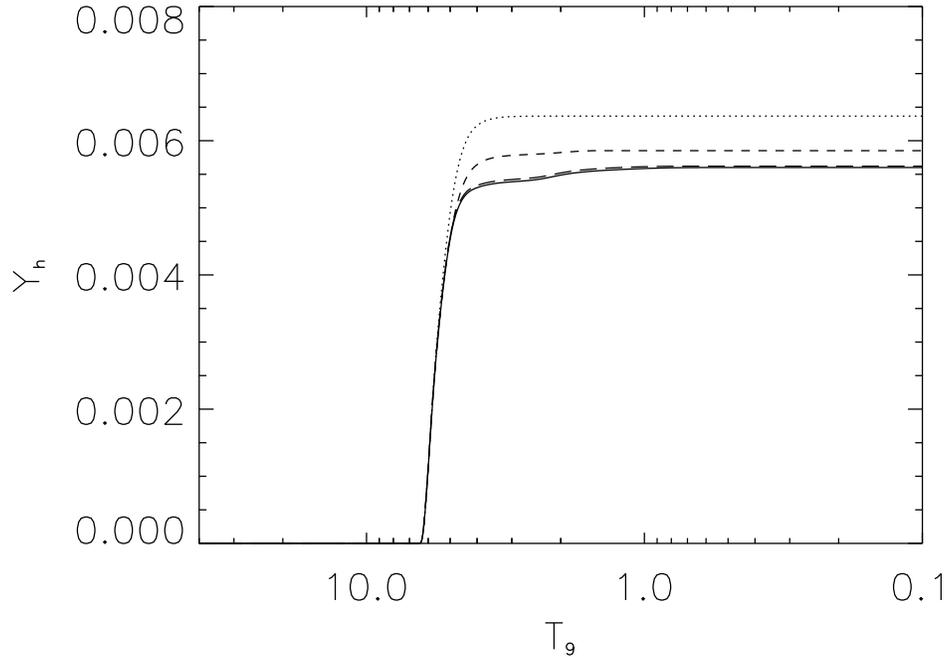,height=4.0in,angle=90}} 
\caption{ 
$Y_h$ during the expansion for models 0 (solid curve), 5 (dotted curve), 6 
(short-dashed curve), and 7 (long-dashed curve).  In models 5, 6, and 7, 
there is only neutrino capture on heavy nuclei for $T_9<7$, $T_9 < 5$, and 
$T_9 < 3$, respectively.  The increased production of alpha particles due 
to neutrino capture on heavy nuclei (see Fig. 15) increases 
the production of heavy nuclei via the three-body reaction sequences 
$\alpha+\alpha +\alpha \to ^{12}$C and $\alpha + \alpha + n \to ^9$Be 
followed by $^9{\rm Be}+\alpha \to ^{12}{\rm C} + n$. 
} 
\label{fig:yhlarge} 
\end{figure}

The QSE-nature of the abundance distributions for $T_9\approx 5$ explains 
the larger final mass fraction of alpha particles in models with
earlier capture on heavy nuclei, as seen in Fig. 
\ref{fig:xalarge}.  As the nuclei capture neutrinos, they become more
proton rich.  In doing so, the average binding of protons in nuclei
decreases.  In the QSE, then, the abundance of free protons tends to
increase.  In practice, the lower proton binding means the average rate of
proton disintegration reactions on heavy nuclei increases somewhat.
The protons, however, are also in equilibrium with neutrons 
and alpha particles; therefore, an increase in the free proton abundance
causes the alpha mass also to increase.  The 
increased alpha mass then leads to further assembly of heavy nuclei via the 
three-body reaction channels.  This is apparent in Fig. \ref{fig:yhlarge}. 
 
\begin{figure} 
\centerline{\psfig{figure=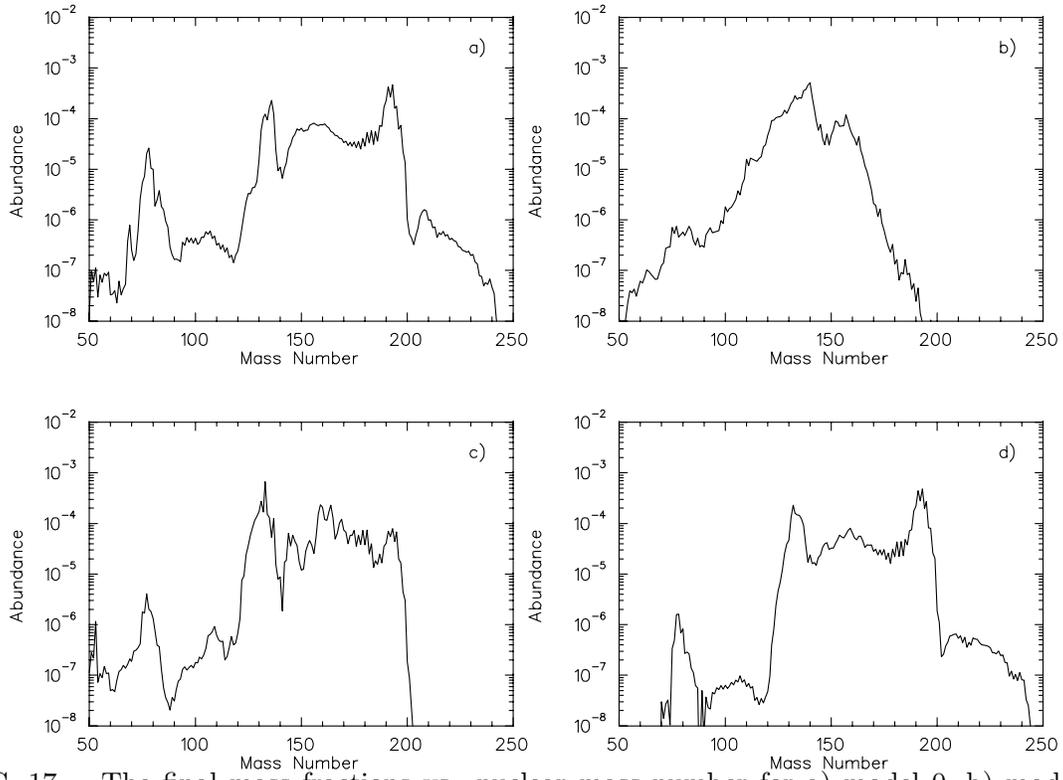,height=4.0in,angle=90}} 
\caption{ 
The final mass fractions vs. nuclear mass number for a) model 0, b) model 5,
c) model 6, and d) model 7. 
Neutrino capture on heavy nuclei during the QSE phase of the expansion has 
hindered the r-process in models 5 and 6.
} 
\label{fig:abund_large} 
\end{figure}

Fig. \ref{fig:abund_large} shows the final abundances versus nuclear mass
for models 0, 5, 6, and 7.  The hindrance of the r-process in models 5 and
6 is readily apparent.  Model 7, however, is little different from model 0,
even though the r-process was accelerated by neutrino captures on nuclei.
As stated previously, it is the neutron-to-seed ratio $R$ that determines
the robustness of an r-process.  Because the neutrino capture on nuclei in
model 7 happens after the large QSE has broken down into 
$(n,\gamma)-(\gamma,n)$
equilibrium clusters, there is little effect on $R$.  The result is that
nuclei capture almost exactly
the same average number of neutrons per nucleus in models 0 and 7.
That the final abundance distributions for these models are so similar
shows that at this flux, neutrino capture is not very 
significant compared to nuclear beta decay during the r-process phase.
 
\section{Implications}     \label{sec:implications}

The essential conclusion of our work is that, given standard neutrino
physics and a realistic neutrino-driven wind, strong neutrino capture
hinders the r-process.  This can happen in several ways.  The largest 
effects come from neutrino capture on free neutrons during the stages
of alpha particle and heavy nucleus formation.  Furthermore, a strong 
neutrino flux will also impede the r-process by inducing significant
neutrino capture  on heavy nuclei
while the material is still in QSE ($T_9 > 4$).  Although neutrino
capture on heavy nuclei after $T_9 < 3$ is not necessarily problematic, 
the simultaneous capture on free neutrons is detrimental. 
In order to examine the effects of charged-current neutrino interactions 
on r-process synthesis, it is necessary to take into account feedback
between nuclear dynamics and weak interactions.  Also, because
neutrino-capture effects at a late stage in the expansion 
imply larger effects at earlier stages in the expansion
(absent neutrino flavor/ type transformation effects), neutrino 
interactions must be self-consistently included everywhere.

The effects we have described pose a severe challenge to the
r-process in stellar explosions.  The r-process components that produced
the solar system's supply of platinum and gold must have taken place
in a sufficiently low flux of normal
electron neutrinos that a strong alpha effect
did not occur.  We conclude that either 1) the r-process occurred in an
environment not associated with a strong $\nu_e$ flux, or 2) in supernovae
fluid elements are carried away extremely rapidly to regions where the 
$\nu_e$ flux is
low, or 3) some non-standard neutrino effects are present.  While we have
employed an exponential outflow model with particular entropy and neutrino
fluxes in this paper, our qualitative conclusions regarding the effects of
charged-current neutrino processes will hold for any general
outflow model.

In fact, however, any attempt to circumvent the harmful influence of the 
alpha effect by invoking convection (or multi-dimensional hydrodynamics)
or general relativity will be problematic.  This stems from  the fact
that it is the weak interaction (changing neutrons to protons and
vice-versa) that is at the heart of the troubles with the r-process 
conditions.  Any model that uses neutrino interactions to supply the
requisite energy to eject nucleons from deep in the gravitational potential 
well of the neutron star will necessarily also have the $Y_e$ of the ejecta 
set by the neutrino capture competition between the processes 
in Eqs. (\ref{eq:ccf})
and (\ref{eq:ccbf}).  This is because a nucleon near the surface of the 
neutron star has a typical binding energy of $\sim 100$ MeV, 
implying that it 
must suffer $\sim 5$ neutrino interactions to be ejected to interstellar
space.  Necessarily, this also implies that the alpha effect and other 
deleterious effects of $\nu_e$ interactions on nucleosynthesis can operate.  
If one
wants the material to be ejected on such a short timescale that $\nu_e$
capture can not ruin r-process 
nucleosynthesis, then one would be forced to find an 
energy source for the ejection process that would not be based on neutrino 
heating.
In turn, this may be difficult to engineer because almost all of the energy
available to the supernova resides in the neutrino seas.  In this case one 
would have to utilize, for example, rotational energy in the core, as in a 
Leblanc-Wilson jet \cite{lw70}.
This would mean giving up on the neutrino-driven wind
model predictions of r-process yields per supernova which naturally agree
with observationally-inferred Galactic heavy element chemical evolution
constraints (e.g. \cite{wh92,mey92}).

Neutrino-driven wind models including general relativistic effects have been
investigated as a method for increasing the neutron-to-seed ratio and 
facilitating the r-process \cite{cardall97,qw97}.  These models have the 
effect of decreasing the timescale and increasing the entropy.  An alpha 
effect will also occur in these models.  The degree to which it impacts
the nucleosynthesis products needs to be investigated.  However, such
general relativistic extensions of neutrino driven wind models suffer
from additional problems such as a requirement for fine tuning of mass 
ejection rates and differential neutrino gravitational redshift.

An alternative idea is that non-standard neutrino physics may allow
r-processing even in a high neutrino flux.  For example, if most electron
neutrinos were converted to other neutrino species (either active or
sterile) by matter-enhanced processes in the region above the neutron
star surface \cite{cqf98,bffm98},
the initial neutron richness would be quite large because
a large flux of anti-neutrinos would drive protons into neutrons. 
Additionally, a reduced $\nu_e$ flux would preclude
reducing this large neutron excess by driving the neutrons back
into protons.  Furthermore, no alpha effect would occur when
nucleosynthesis begins.  Many scenarios of this type exist and merit
investigation, although we again caution one must include all neutrino
effects and must properly take the nuclear dynamics into account.

The strong sensitivity of r-process yields to the neutrino flux present
interesting implications for the extinct r-process radioactivities.  Chief
among these isotopes are $^{129}$I and $^{182}$Hf, with half-lifes of 16
Myr and 9 Myr, respectively.  Meteoritical evidence indicates that these
isotopes were alive in the early solar system, an observation that
apparently constrains galactic nucleosynthesis over the last several
million years prior to collapse of the solar nebula.  The curious aspect of
these isotopes is that the inferred live $^{182}$Hf abundance in the early
solar system roughly agrees with expectations from continuous Galactic
nucleosynthesis while the inferred  $^{129}$I abundance fails to reach
the same expectations by about a factor of 100.  An obvious
explanation is that there are different kinds of supernovae which occur
with different frequencies \cite{wass96,qwg97}.  

In any case, perhaps the clues about the r-process synthesis rates in the
Galaxy provided by considering $^{129}$I and $^{182}$Hf could give us
insights into how nature manages to circumvent the alpha effect.  It
could be that the supernova events that are responsible for the production
of $^{129}$I represent only a \lq\lq partial\rq\rq\ r-process
where vigorous neutron capture to or
beyond mass 130 is hindered by the alpha effect.
Likewise the $^{182}$Hf production events have somehow managed to completely
disable the alpha effect.

In summary, strong neutrino capture during expansions of neutron-rich matter
greatly hinders production of r-process isotopes.  It will be fascinating
to see how this dramatic conclusion will lead to new insights into supernova
dynamics, neutrino physics, and the r-process of nucleosynthesis.

\acknowledgments

This work was supported in part by NASA grant NAGW-3480 at Clemson and NSF
grant PHY95-03384 at UC San Diego.

\begin{table}
\squeezetable
\caption{
Parameters for our eight model calculations.  For all
calculations, the entropy was a constant 110 $k$ per nucleon and
$T_{\nu_e} = 3.5$ MeV and $T_{\bar{\nu}_e} = 4.0$ MeV.
}
\begin{tabular}{llllll}
\\
Model\ \ \  & $L_{\nu_e}$ (ergs/s)\ \ \ & $L_{\bar{\nu}_e}$ (ergs/s)
\ \ \ & $\nu_e , \bar{\nu}_e$-capture on free nucleons\ \ \ &
$\nu_e ,\bar{\nu}_e$-capture on nuclei \ \ \ &  Other comments \\
\\
\hline \\
0 & $10^{51}$ & $4 \times 10^{51}$ & for $T_9 \geq 10$ & No & Reference
calculation \\
\\
1 & $10^{51}$ & $4 \times 10^{51}$ & for $T_9 \geq 10$ & for $T_9 \geq 10$
and $T_9 \leq 3$ & ------- \\
\\
2 & $10^{51}$ & $4 \times 10^{51}$ & for $T_9 \geq 10$ and $T_9 \leq 3$ & 
for $T_9 \geq 10$ and $T_9 \leq 3$ & ------- \\
\\
3 & $10^{51}$ & $4 \times 10^{51}$ & On throughout & On throughout & Most
realistic \\
\\
4 & $10^{51}$ & $4 \times 10^{51}$ & for $T_9 \geq 10$ and $T_9 \leq 3$
& for $T_9 \geq 10$ and $T_9 \leq 3$ & No $^3{\rm H}(\alpha,\gamma)^7{\rm Li}$\\
\\
5 & $10^{52}$ & $4 \times 10^{52}$ & for $T_9 \geq 10$ & for $T_9 \geq 10$
and $T_9 \leq 7$ & ------- \\
\\
6 & $10^{52}$ & $4 \times 10^{52}$ & for $T_9 \geq 10$ & for $T_9 \geq 10$
and $T_9 \leq 5$ & ------- \\
\\
7 & $10^{52}$ & $4 \times 10^{52}$ & for $T_9 \geq 10$ & for $T_9 \geq 10$
and $T_9 \leq 3$ & ------- \\
\\
\end{tabular}
\end{table}

\clearpage

\clearpage
 
\clearpage
 
\clearpage
 
\clearpage
 
\clearpage
 
\clearpage
 
\clearpage
 
\clearpage
 
\clearpage
 
\clearpage
 
\clearpage
 
\clearpage
 
\clearpage

\clearpage
 
\clearpage
 
\clearpage
 
 
\end{document}